\documentclass[iop,apj]{emulateapj}
\usepackage{amsmath}
\begin{document}
\title{MHD Simulations of AGN Jets in a Dynamic Galaxy Cluster Medium}
\author{P. J. Mendygral\altaffilmark{1,2,3}}
\author{T. W. Jones\altaffilmark{1,2}}
\author{K. Dolag\altaffilmark{4,5}}
\altaffiltext{1}{School of Physics and  Astronomy, University of Minnesota, Minneapolis, MN 55455}
\altaffiltext{2}{Minnesota Supercomputing Institute, University of Minnesota, Minneapolis, MN 55455}
\altaffiltext{3}{Cray Inc., 380 Jackson Street, Suite 210, St. Paul, MN 55101}
\altaffiltext{4}{University Observatory Munich, Scheinerstr. 1, D-81679 Munich, Germany}
\altaffiltext{5}{Max Planck Institute for Astrophysics, Karl-Schwarzschild-Str. 1, D-85741 Garching, Germany}


\begin{abstract}
We present a pair of 3-d magnetohydrodynamical simulations of intermittent jets from a central active galactic nucleus (AGN) in a galaxy cluster extracted from
a high resolution cosmological simulation.  The selected cluster was chosen as an apparently relatively relaxed 
system, not having undergone a major merger in almost 7 Gyr. 
Despite this characterization and history, the intra-cluster medium (ICM) contains 
quite active ``weather''. We explore the effects of  this ICM weather
on the morphological evolution of the AGN jets and lobes.  The orientation of the jets is different in the two  simulations
so that they probe different aspects of the ICM structure and dynamics.  We find 
that even for this cluster that can be characterized as
relaxed by an observational standard, the large-scale, bulk ICM motions can 
significantly distort the jets and lobes.  Synthetic X-ray observations of the
simulations show that the jets produce complex cavity systems, while synthetic 
radio observations reveal bending of the jets and lobes similar 
to wide-angle tail (WAT) radio sources.  The jets are cycled on and off with a 26 Myr period using a 50\%
duty cycle. This leads to morphological features similar to those in ``double-double'' radio galaxies. 
While the jet and ICM magnetic fields are generally too weak in the simulations
to play a major role in the dynamics, Maxwell stresses can still become
locally significant.
\end{abstract}


\section{Introduction}
\label{introduction}
Observations of powerful outflows from active galactic nuclei (AGN) show that they have a considerable effect on the energy budget and morphology
of their host galaxy cluster.  
The jets from the super-massive black hole that powers an AGN inflate low density bubbles in the host galaxy cluster's intra-cluster medium (ICM) 
that are observed as X-ray cavities.  The minimum energy required to inflate these cavities can be as high as 
10$^{61}$ erg \citep[\emph{e.g.,}][]{mcnamara05,wise07} and typically imply a power on the 
order of 10$^{41}$ - 10$^{44}$ erg s$^{-1}$ averaged over the age of the cavity \citep[\emph{e.g.,}][]{birzan04,mcnamara07,birzan08,diehl08}.
Most X-ray cavities are paired as a result of the twin AGN jets that formed them, and most are filled with radio synchrotron emission
from relativistic, cosmic ray electrons (CRs); that is, ``radio lobes.'' 
Absent thermal conduction, models of the ICM without any heating source predict that galaxy clusters 
should establish ``cooling-flows", where radiatively cooled ICM plasma falls towards the central galaxy driving up star formation to very high values \citep{fabian94}.  
Observations do not show evidence of these classical cooling-flows, but in contrast  they show that mass accretion onto the central galaxy is at least an order
of magnitude below cooling-flow predictions \citep{peterson06}.  One popular explanation for regulating cooling in clusters  is that energy output
from the AGN in the central galaxy heats the ICM sufficiently to limit accretion. Indeed, the  enthalpy content estimated in X-ray cavities 
appears to be sufficient to offset cooling of the ICM \citep[\emph{e.g.,}][]{birzan04,birzan08}.  While the details of the physical 
processes involved in the transfer of energy from AGN outflows to the ICM are not fully understood, it is generally agreed that these outflows 
contribute to the dynamics and thermodynamics of the ICM.

At the same time there is also good evidence that the ICMs significantly affect AGN outflows.
For example, several radio lobes that have been observed, such as Hydra A \citep{wise07} and 
3C 31 \citep{laing08}, show complex morphologies with various kinks and turns.  The cited authors suggested that, for these objects, the jets and lobes 
may be reacting to bulk motions of the ICM\footnote{In this work we use "bulk flow" or "wind" to mean motions of the ICM that are
not characterized by isotropic turbulence. The complete motion of the
ICM is described by the combination of bulk flow and turbulence on scales smaller than the cluster core.}.  Further, there are groups of AGN-produced radio sources specifically classified by their bent morphologies; namely, 
narrow-angle tails (NATs) and wide-angle tails (WATs), depending on the
apparent degree of bending.  \citet{owen76} proposed that these objects obtain their bent shape from the relative motion through the ICM
of the galaxy hosting the AGN engine.  In the case of WATs, the low peculiar velocity of the associated galaxies in combination with
reasonable assumptions for the density and
velocity of the plasma within the AGN outflows suggest that the motion of the ICM 
within the cluster is more responsible 
for their characteristic C-shape than the motion of the
individual host galaxy within the cluster \citep[\emph{e.g.,}][]{burns94,douglass08}.
\citet{burns94} argued that cluster mergers can produce sufficient 
ram pressure in the ICM to cause this shape, and simulations have shown that motions produced by merger events with sufficient ram pressure to bend WATs are 
persistent for timescales exceeding 10$^{8}$ years \citep{loken95,roettiger96,ricker01}.  
Similarly, recent simulations show large-scale ICM ``sloshing'' several
gigayears after a gravitational encounter with a passing subcluster \citep[e.g., ][]{zuhone10}.

Direct detection of such bulk ICM flows became 
feasible with the \emph{Advanced Satellite for Cosmology and Astrophysics} (ASCA), and
the \emph{Chandra X-ray Observatory} has further improved such measurements.  \citet{dupke06}, for example, used \emph{Chandra} observations to 
explore the velocity structures in the Centaurus Cluster (Abell 3526), and they found that significant departures from hydrostatic equilibrium
are possible even in a cool cluster that shows little evidence of a strong merger.  ICM sloshing motions have also been observed 
in the central regions of several cool-core clusters
\citep[see][for a review]{markevitch07}.

There is limited simulation literature addressing interactions between AGN outflows and dynamical
ICM flows, or  ``weather.''
For example, \citet{bals92} and \citet{porter09} carried out 3d simulations of AGN outflows
into supersonic cross-winds that produced NAT-like structures. 
\citet{loken95} used 3-d hydrodynamic simulations of a low density jet obliquely  crossing a plane shock to demonstrate that merger produced 
shocks can decollimate a jet and cause it to bend.  In a study of AGN energy deposition to the ICM \citet{heinz06} and \citet{morsony10} 
used output from a cosmological simulation as initial
conditions for hydrodynamical simulations that included a pair of AGN jets. To differentiate behaviors fully from hydrostatic clusters 
the latter study specifically selected a cluster that was far from being relaxed and that included a net rotation and a dynamically induced cold front.  
Not surprisingly those authors found significant effects on jet propagation and lobe 
morphology due to the rather severe ICM weather.

Especially in the context of clusters exhibiting cavities produced by central dominant galaxies, which more typically associate with apparently more
relaxed clusters, however, it is important to understand if such effects can take place
in those types of clusters.  Except for the \citet{porter09} idealized NAT study, there are no published simulations of AGN interactions with dynamical ICMs that included magnetic fields.  However, since magnetic fields both
introduce potential dynamical influences and are also essential
to modeling the generation and transport of diagnostic nonthermal emissions,  it is important to extend such simulations into the MHD regime.
\citet{oj} (hereafter OJ10) recently conducted 
a series of 3-d MHD simulations of steady and intermittent jets 
in an analytically defined, magnetized ICM close to hydrostatic equilibrium. The magnetic field was disordered on scales similar to the size of the ICM core
and on average produced a pressure roughly 1\% of the gas pressure, so probably characteristic of cluster strength fields. 
They found that for localized regions, those magnetic fields can have a significant effect on
the morphology of the jet lobes.  Whether this remains true for a dynamically self-consistent cluster evolved through a cosmological simulation has not been examined.
Another important result from the OJ10 simulations was the lack of any significant deflections of the jets or lobes.  The steady and intermittent jets in their study both maintained relatively
straight-line paths, suggesting that a dynamical ICM is required to produce bent jet and/or bent lobe morphologies. 

To address some of these open issues we present here a pair of 3-d simulations of MHD jets in a dynamic, magnetized ICM extracted from a high resolution cosmological simulation.  The selected cluster
was chosen from among the more relaxed systems in the simulation.  To expand the insights into these interactions and the influence of ICM weather
on radio lobe morphology the two simulations directed the AGN jet pairs in two mutually
orthogonal directions that probed distinct regions of the ICM.  
The paper is organized as follows:
\S \ref{setup} describes the numerical setup for the calculations, \S \ref{cluster_cond} describes the initial conditions in the cluster,
\S \ref{sim_results} qualitatively describes the results of the simulations, \S \ref{weather} discusses the effects of weather on the jets and lobes,
and \S \ref{weather_conclusions} summarizes the major conclusions. We assume below that $H_{0}=70$ km s$^{-1}$ Mpc$^{-1}$, 
$\Omega_{M}=0.3$, and $\Omega_{\Lambda}=0.7$.


\section{Numerical Details}
\label{setup}
The simulations of AGN jets for this study were performed on a 1008x1008x1008 three dimensional Cartesian grid with a uniform resolution of 
$\Delta x = \Delta y = \Delta z = 1$ kpc per zone using WOMBAT \citep{mend11b}, which is a versatile,  efficient and highly scalable MHD
code.  Here, WOMBAT evolved the equations for non-relativistic ideal MHD with adiabatic index, $\gamma = 5/3$,  using the second order, Total Variation Diminishing (MHDTVD) method 
of \citet{rj95}.
 The solenoidal constraint of the magnetic
field was maintained to machine accuracy with the ``constrained transport'' (CT) method of \citet{rj98}, which solves the induction
equation with a directionally unsplit update using EMFs derived from the Riemann solver. 
The simulations
included the total energy conserving method for gravity available in WOMBAT and described in detail in \citet{mend11c}.
This method adds gravitational source terms to MHDTVD in a manner that maintains 2$^{\text{nd}}$ order code accuracy.
The evolution of a passive
cosmic ray electron (CR) momentum  distribution, $f(\vec x, p, t)$, was also calculated in the simulations with a restricted form of the ``Coarse Grained Momentum Volume'' (CGMV) \citep{jk05}
implementation in WOMBAT.  CGMV solves the diffusion-convection equation for a population of relativistic particles. For these
simulations spatial CR diffusion outside shocks was neglected, while adiabatic changes in momentum, test-particle diffusive acceleration at shocks and radiative losses by synchrotron emission and inverse Compton scattering of 
Cosmic Microwave Background (CMB) photons were included (see further details below and, e.g., \citet{treg01}).  

The boundary conditions for these simulations were based on a technique used by OJ10.  To
limit any influence from the boundaries, we used a very large domain (over a Mpc along a side) and modified continuous boundaries
that limit the fluxes through those surfaces.  In the boundary zones a hydrostatic atmosphere with a constant sound speed
was derived by extrapolating conditions from the last physical zone.  Since the mass distribution was known on a much larger domain (see below), the gravitational potential
was known out to radii far beyond the computational domain. Consequently, the exact value for the potential described below was used through the boundaries. 

\subsection{Setting up the Cluster}
\label{cluster_setup}
The initial conditions for the simulations described here were extracted from a high resolution simulation of a galaxy
cluster performed with an MHD implementation of the smoothed particle hydrodynamic (SPH) code GADGET-3 \citep{mhd_gadget}.
The selected cluster is \emph{g676} from \citet{dolag09}, resimulated at extreme high resolution with MHD enabled 
(for details see Stasyszyn \& Dolag, in prep). In short, the cluster used for this study has been extracted from a re-simulation
of a Lagrangian region selected from a cosmological, lower resolution DM-only simulation \citep{2001MNRAS.328..669Y}. This parent 
simulation has a box--size of $684$Mpc and assumed a baryon fraction of  $f_{\rm bar}=0.13$ and $\sigma_8=0.9$ for the normalization 
of the power spectrum. Using the ``Zoomed Initial Conditions'' (ZIC) technique \citep{tormen97}, this region was re-simulated with 
higher mass and force resolution by populating the Lagrangian volume with a larger number of particles, while appropriately adding
additional high--frequency modes drawn from the same power spectrum. The initial unperturbed particle distribution (before 
imprinting the Zeldovich displacements) was realized through a relaxed glass-like configuration \citep{1996clss.conf..349W}.
Gas was then added to the high-resolution regions by splitting each parent particle into a gas and a DM particle. The gas and the
DM particles were displaced by half the original mean inter-particle distance, such that the centre-of-mass and the
momentum of the original particle are conserved. The final mass resolution of the dark matter and gas particles in our simulations
is $m_{\rm dm}=12.1\times 10^6\,{\rm M}_\odot$ and $m_{\rm gas}=2.2\times 10^6\,{\rm M}_\odot$ respectively. Thus, the clusters 
within its virial radius is resolved with $11\times10^6$ dark matter and a similar number of gas particles. In the simulations, 
the gravitational softening length was chosen to be $\epsilon=1.4\,\mathrm{kpc}$ which correspond roughly to the mean
particle separation in the center of the cluster at z=0. For the initial magnetic field we choose a space filling, homogeneous,
primordial magnetic field of $B=10^{-11}$G co-moving as in \citet{dolag09}.

\begin{figure*}
\vspace{1.0in}
\includegraphics[height=.4\textheight]{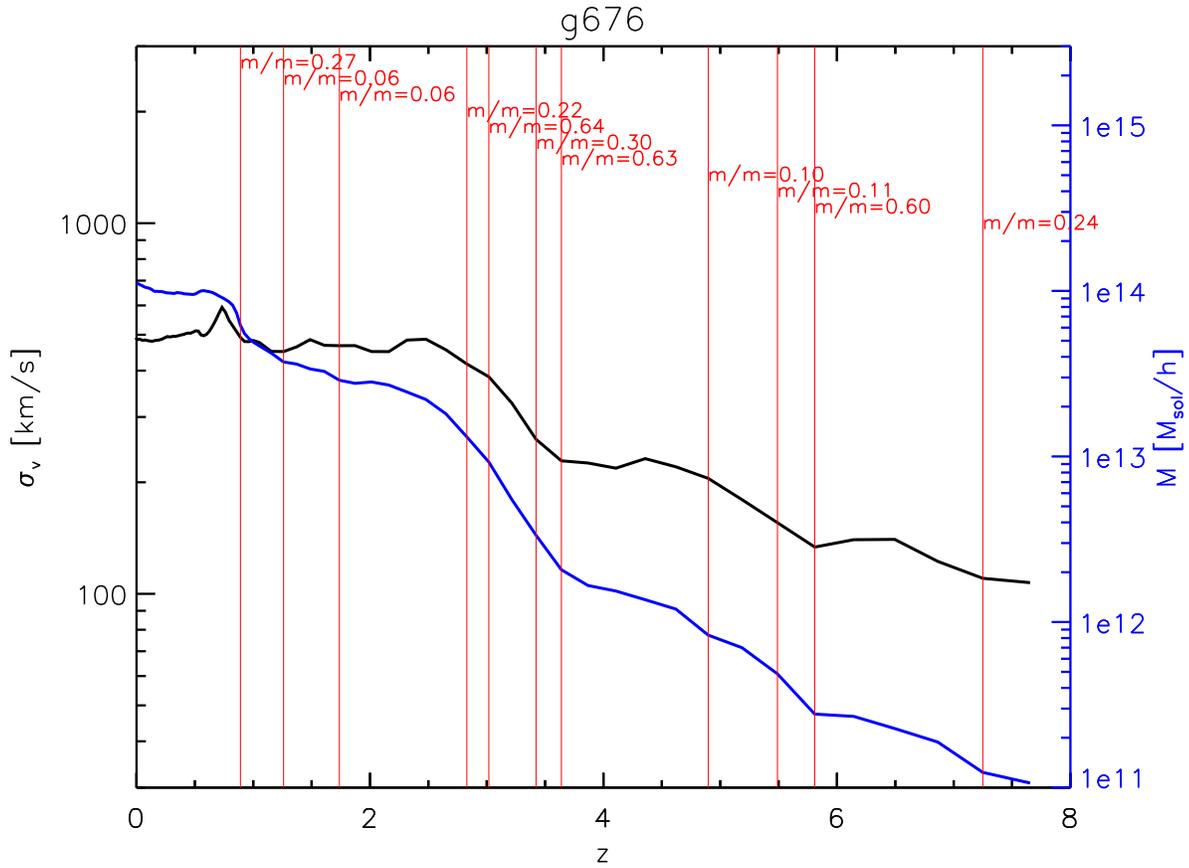}
\caption[Fig 1 Velocity dispersion and dark matter mass with redshift for the cluster.]{Evolution of velocity dispersion (dark matter)
and total mass as a function of redshift for the selected galaxy cluster \emph{g676}. The vertical lines mark identified merger events with a mass
ratio indicated by the labels.}
\label{fig:cluster_mass}
\end{figure*}

For this work, we were interested in assessing the effects of cluster ``weather'' specifically for a cluster that observationally appears relaxed yet harbors significant bulk ICM flows.
We selected the state of \emph{g676} at a redshift of $z \sim$ 0, which was the least dynamic period for this cluster.  
As seen in Figure \ref{fig:cluster_mass}, the last major merger (mass ratio of 0.27) occurred at $z \approx$ 0.8 (a little 
less than 7 Gyr before our jet simulations start), and the dark matter mass
continued to grow slowly after that.  The general morphology and environment of the cluster  at z = 0 can be seen in Figure 1 of \citet{dolag09}.
Volume averaged values for baryonic density, temperature, velocity, and the magnetic field were all computed from the SPH particles 
and mapped to a uniform grid matching the jet simulation grid.  The simulation assumed a hydrogen fraction of $X = 0.76$ and a helium fraction of $Y = 0.24$, resulting in
a mean molecular weight of $\mu = 0.59$.  An ideal equation of state was used to convert between pressure and temperature.  Radiative cooling
of the thermal plasma was not included in the GADGET calculations that formed
cluster \emph{g676}, so neither was it used in the jet simulations.
A summary of relevant cluster ICM properties is given in \S \ref{cluster_cond}.

\begin{figure*}
\centerline{\includegraphics[height=.37\textheight]{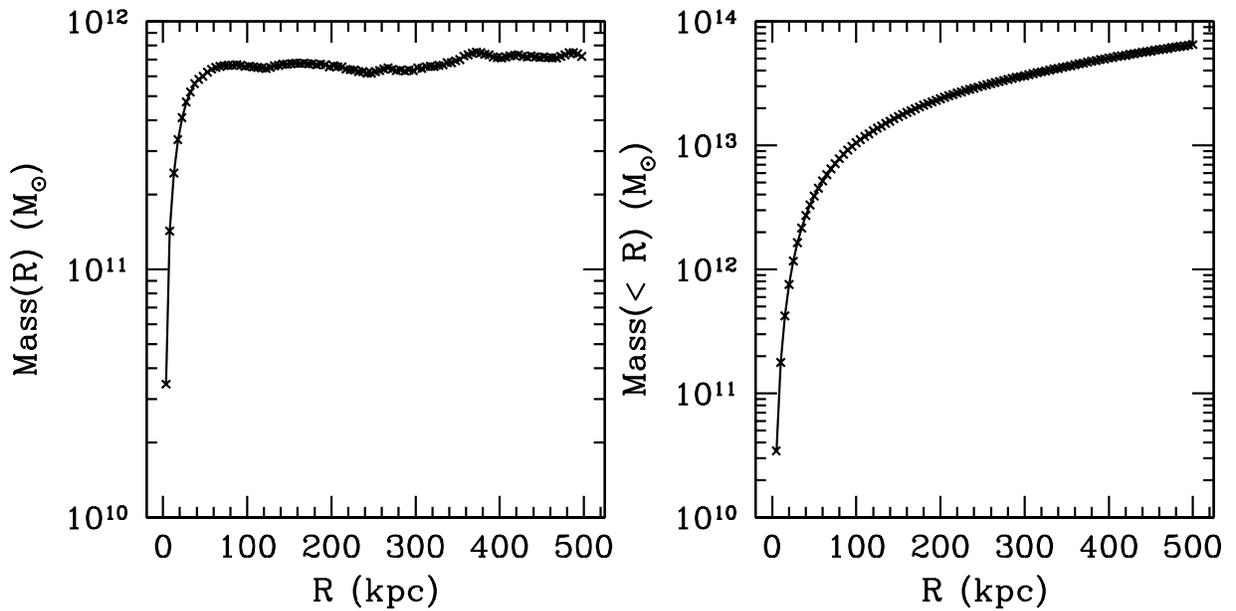}}
\caption[Fig 2 Gravitating mass profile for the cluster.]{Gravitating mass profile used at all times to define the gravitational potential.
The mass includes both dark matter and
baryonic matter from the GADGET simulation and was derived from an azimuthal average at each radius.}
\label{fig:cluster_gravmass}
\end{figure*}

For the jet simulations, we used a static gravitational potential defined from the total gravitating mass 
(dark matter and baryonic matter)
azimuthally averaged from the SPH values at the time used for the initial conditions.  Figure \ref{fig:cluster_gravmass} shows the gravitating
mass as a function of radius.  The total enclosed mass out to the virial radius of 1.4 Mpc is 
$1.53 \times 10^{14} \text{M}_{\odot}$.  The decision to use a static potential as opposed to evolving a population of DM 
particles and maintaining an updated full potential was based on simplicity, the relaxed nature of the cluster and the relatively brief duration of the jet activity (200 Myr) compared to the dynamical timescale for the cluster mass distribution.
Obviously over longer times or in a cluster involved in significant active restructuring  it would be important to follow the evolution of the gravitational potential.

\subsection{Setting up the Jets}
\label{jet_setup}

For each simulation, a pair of oppositely directed jets was launched from a cylinder centered in the grid ($X = Y = Z = 0$).  The launching cylinder had a radius of
$r_{j} =$ 3 kpc and length on each side of the jet origin of $l_{j}$ = 8 kpc.  An additional 3 kpc thick collar and cap surrounded the cylinder to transition from
jet conditions to ICM conditions and to contain the return electric current associated with the jet magnetic field (\S \ref{jet_field}). 
Special consideration was given to
how the cylinder interacted with the ICM.  In these simulations, ambient flows could in principle enter the cylinder, even from the initial conditions.
This can be particularly problematic if ICM magnetic fields enter the cylinder, where they can blend with the jet field and produce undesirable 
field geometries.  

Two steps were taken 
to alleviate this. First, at the start of the simulation the ICM material in the vicinity of the jet cylinder location was gently pushed aside 
 before the jet cylinder was initialized. In addition,  reflecting boundaries were defined as defaults for the cylinder.  
The initial ICM evacuation was accomplished through a purely solenoidal velocity field imposed at the start
of the simulation on a sphere with radius $r_{j}$ centered on the grid with $v_{\phi} = v_{e}(r/r_{j})$ , and $v_{e}$ set to the local sound speed.  This sphere was over-pressured by 90\% as well.  These conditions were allowed to evolve for 3.28 Myr,
which was the sound crossing time over $r_{j}$.  This removed $\approx$ 25\% of the ICM plasma and magnetic energy from the
jet cylinder region.
The jet cylinder was then established at $t = 3.28$ Myr. To divorce the ICM and jet magnetic fields, any remaining ICM
field penetrating the jet cylinder or collar was set to zero at this time.  This last step effectively introduced some magnetic monopoles around the cylinder, but they
had negligible effect on the jet flows and were maintained to low levels by the CT scheme in WOMBAT.  
The collar region around the active  jet cylinder acted as an impenetrable electrically conducting surface because of the
imposed reflecting boundaries for both the normal momentum and magnetic field.  This prevented loss of any conserved quantity into the launching 
region.  The caps of the cylinder transitioned between reflecting boundaries and  outflowing jet properties as the jet cylinder transitioned between active and inactive states.

An important feature of the jet cylinder for this investigation was that it could be
oriented in any direction on the grid.  This allowed jets to be launched at an arbitrary angle with respect
to the large scale flows and pressure structures in the ICM.  To achieve this, we defined all vector quantities in the jet cylinder according to 
cylindrical coordinates.  The jet cylinder is azimuthally symmetric, so only two coordinates are needed, $z$ and $r$, where $z$
is along the cylinder symmetry axis and $r$ is radially outward.  By default the jet $z$ axis was aligned with the grid $Z$ axis,
and Euler angles were used to reorient the cylinder.  After the rotation, all cylinder coordinates were converted into grid coordinates
so that quantities could be mapped to the Cartesian grid. The specific orientations
of the two jets in this study are discussed in section \ref{sim_results}.

The active jet density and pressure were set to 
$\rho_{j} = 4.165\times 10^{-28}~{\rm g~cm^{-3}}$ and
$P_{j} = 2.499\times 10^{-10}~{\rm dyne~cm^{-2}}$ (approximately 1/200 the initial local ICM density) and in approximate pressure equilibrium 
with the initial surrounding ICM.  For the assumed adiabatic index of $\gamma = 5/3$, this gave a sound speed inside the jet of $c_{s,j} = 10^{4}$ km s$^{-1}$
or 0.03c.  A passively advected ``color" (or mass fraction) tracer, $C_{j}$, was also introduced in the jet cylinder that was used to identify AGN and ICM plasma.  The color was
set to 1 in the jet cylinder, while all ICM material was initially assigned $C_{j} = 0$.

\subsubsection{ Jet Velocities}
When it was active the jet velocity in the jet cylinder was given a spatial ramp from its origin in $z$ to allow a smooth transition between the oppositely directed flows of the two jets.  
In particular, the velocity had the form, $v_{z,j}(z) =v_j \mathcal{ F}(z)$, where
\begin{align}
\mathcal{F}(z) = \text{TANH}\left(1.8 \left[\frac{z}{l_{j}}\right]^{2}\right). \label{f:vjet}
\end{align}
The value 1.8 was determined empirically by minimizing the automatic application of diffusive, positive-definite HLL fluxes by WOMBAT 
in the jet cylinder, which are used to eliminate unphysical pressures during peak flow times. 
The velocity in the jet cylinder was defined with this ramp 
to be $v_{z,j}(z,t) = v_{j}(t)F(z)$, where $v_j(t) = 1.2 \times 10^{4}$
km s$^{-1}$, when the jet was fully on (see \S \ref{jet_evolve}).  This yields a pair of supersonic emergent jets with the flow at Mach number, $M_j = 1.2$, with respect to the jet sound
speed.  The peak speed of $v_{j}$ corresponds to a Lorentz factor of 1.0008 and is sufficiently low that a non-relativistic
MHD solver is applicable.  The jet velocity was uniform in $r$ across the jet core ($r\le 3~\rm{kpc}$), with a transition to the ambient medium velocity $\propto r^{-1/3}$.

The core luminosity of each of the two jets, ignoring small contributions from magnetic energy, is given by
\begin{align}
L_{j} &= \pi r_{j}^{2} v_{j}\left(\frac{1}{2}\rho_{j}v_{j}^{2} + \frac{\gamma}{\gamma - 1}P_{j}\right) \nonumber
\\ &= \pi r_{j}^{2}\frac{1}{2}\rho v_{j}^3\left(1 + \frac{2}{M_j^2(\gamma - 1)}\right) ,
\end{align}
where $\gamma = 5/3$.  When the jets are fully active, the preceding parameters give a total luminosity (including both jets) of 
$L_{j} = 6\times 10^{44}$ erg s$^{-1}$.  The transitional collar region adds roughly another 50\% to the actual jet
luminosity. The ratio of kinetic energy flux to enthalpy flux for these jets
was $\rho_{j}v_{j}^{2} / 5P_{j}$ = 0.48.  The actual total energy injected onto the grid by the jet cylinder in these runs was
$2.2 \times 10^{60}$ erg.

For later discussion it is also useful to know the thrust of the jets, $F_j$. The thrust is easily computed from the luminosity (with $\gamma = 5/3$) as
\begin{equation}
F_j =L_j \frac{2}{v_j (1 + 3/M_j^2)},
\end{equation}
so that the individual core jet thrusts at peak luminosity are $F_j \approx 1.6\times 10^{35}$ dyne.

\subsubsection{ Jet Magnetic Field}
\label{jet_field}
The magnetic field in the jet launching cylinder was derived with a method that maintained $\nabla \cdot \mathbf{B}$ to machine accuracy.  
The jet field was purely toroidal inside the jet cylinder
with $B_{\phi} = B_{j}(r/r_{j})$ when $r < r_{j}$, where $B_{j} =$ 7.92 $\mu$G (plasma $\beta = 8 \pi P_{j}/B_{j}^{2}$ = 100).  
Outside of that radius, the
field dropped off as $r^{-3}$ and was zero outside of the jet cylinder collar.  This kept the jet field separated from
the ICM field and made the net electric current in each jet zero.  The arbitrary orientation of the jet cylinder was an additional complication for defining the jet field.
We implemented this by defining an electric field that determined the rate of change of the magnetic field in the cylinder according to Faraday's 
Law, $-\partial\mathbf{B}/\partial t = \nabla \times \mathbf{E}$.  Solving this equation allowed magnetic field to be replaced 
as it was advected out of a zone at an arbitrary angle.  Given the desired form of the magnetic field and the jet velocity as a function $z$, 
this electric field was
\begin{align}
E_{z}(r,z) = \begin{cases}
                 \scriptstyle B_{j}v_{j}(z)\frac{r_{j}}{2\Delta z}\left\{\left(\frac{r}{r_{j}}\right)^{2} - \left[\frac{5}{4} - 
                 \frac{1}{4}\left(\frac{r}{r_{j}}\right)^{4}\right]\right\} &
                 \\ \text{\ \ \ \ for } r < r_{j} \text{ and } |z| \le l_{j} \\
                 \scriptstyle B_{j}v_{j}(z)\frac{r_{j}}{2\Delta z}\left\{\left[\frac{5}{4} - 
                 \frac{1}{4}\left(\frac{r_{j}}{r}\right)^{4}\right] - 
                 \left[\frac{5}{4} - \frac{1}{4}\left(\frac{r_{j}}{r_{b}}\right)^{4}\right]\right\} &
                 \\ \text{\ \ \ \ for } r_{j} < r < r_{b} \text{ and } |z| \le l_{j} \\
                 \scriptstyle 0 & \\ \text{\ \ \ \ otherwise},
           \end{cases}
\end{align}
where $r_{b} = r_{j} + 3\Delta z$.  The magnetic field added to the zones in the jet cylinder and collar was
simply $\delta \mathbf{B} = \Delta t(\nabla \times \mathbf{E})$, where $\Delta t$ is the current time step.  The value of 
$\nabla \times \mathbf{E}$ only needed to be computed once at the beginning of the simulation.

\subsubsection{Jet Source Time Evolution}
\label{jet_evolve}
AGN activity is likely to be unsteady on the $\ge 10^8~\rm{yr}$  timescales required to form large scale cavities, while the intermittency of jet activity has 
significant impact on the interactions with the ICM  (see, e.g., OJ10). In these simulations the activity of the jets was toggled on and off
over approximately 26 Myr as a simple way to mimic such cyclic activity.  In particular we applied to the characteristic jet velocty, $v_j$,  and magnetic field, $B_j$, a ramp function, $R(t)$, defined as
\begin{align}
R(t) = \eta(t) \text{MAX}\left[0,\text{TANH}\left(\frac{2(t - t_{\text{on}})}{t_{\text{blend}}}\right)\right] + \\ \nonumber
       \left(1 - \eta(t)\right)\text{MAX}\left[0,-\text{TANH}\left(\frac{2(t - t_{\text{off}})}{t_{\text{blend}}}\right)\right], \\
\eta(t) = \begin{cases}
                       0 & \text{for } t \ge \frac{t_{on} + t_{off}}{2} \\
                       1 & \text{otherwise},
          \end{cases}
\end{align}
where $t_{\text{on}}$ and $t_{\text{off}}$ are the start and end time for the current jet cycle and $t_{\text{blend}} = 3.28$ Myr.  
A duty cycle of $\sim$ 50\% was used for both 
simulations with a period of 26.2 Myr.  Thus, the jet velocity and added magnetic field were given a time dependence as
$v_{j}(z,t) = v_{j}(z)R(t)$ and $\delta \mathbf{B}(r,z,t) = \Delta t[\nabla \times \mathbf{E}(r,z)]R(t)$.


\subsection{Cosmic Ray Electrons}
\label{cr_setup}

In order to enable realistic synthetic observations of nonthermal emissions we incorporated in the jet outflows a population of test particle CR 
electrons with Lorentz factors from 10 to $1.6\times 10^5$ and evolved their distribution, $f(\vec r, p, t)$, using the CGMV routine in WOMBAT.
This range in Lorentz factors is sufficiently large for the typical magnetic field strengths in these simulations to model
synchrotron emission from a few MHz to tens of GHz. No CR electrons were included in the initial ICM, 
although any electrons introduced by the jets and then mixed into the ICM were followed
in the CGMV scheme.  For these simulations, CRs were subject to adiabatic changes in momentum as
well as energy losses due to synchrotron emission and inverse Compton scattering of CMB photons.  CRs were also subject to 1$^{\text{st}}$
order Fermi acceleration  at shocks (diffusive shock acceleration (DSA)).  

As described by \citet{treg01}, the characteristic scattering length for electrons 
in this energy range and in field strengths of a few $\mu$G, typical for these simulations, is on the scale of the solar system, and
the acceleration timescale is on the order of years.  These scales are both unresolved in these simulations, which resolves
lengths of a kiloparsec and time in kiloyears.  This mismatch in scales can be exploited by assuming that the acceleration process
for CRs at a shock is completed in a time that is far less then a typical dynamical time step in the simulation.  For this application,
the CGMV implementation does not include diffusion terms and simply redistributes existing post-shock CRs into a power law according to
test particle DSA theory.  The slope $q_{s}$ of the resulting power law is related to the measured 
compression ratio of the shock, $r$, by $q_{s} = \frac{3r}{r-1}$. To implement DSA WOMBAT includes a directionally unsplit shock detector
that locates and characterizes shock strengths. Eight boundary zones along domain boundaries allowed accurate shock detection
and characterization up to the edge of the local domain boundary and the physical grid boundary.

The CR population was introduced in the flow of AGN plasma out the jet cylinder. Since the CR electrons were passive, their number was
somewhat arbitrary, except in the calculation of nonthermal emissions. For the synthetic observations reported here, we
assume the emergent jets carried a population of CRs with a density of $5\times 10^{-7}$ cm$^{-3}$
(about 0.2\% of the jet bulk proton number density) and had a single
power law slope of $q_{j}$ = 4.5. For consistency this population should not be expected to contribute significantly to the subsequent flow
dynamics. In fact, with this density and slope, the nominal CR electron pressure at the jet source was
 $4\times 10^{-12}$ dyne cm$^{-2}$, or approximately 2\% of the ``thermal'' gas pressure in the AGN plasma. Especially, given the softer equation of state
for these CRs, the downstream CR pressure influences would, indeed, have been minor. 
Since the nonthermal emissions of interest to us here are associated with the jets,  we disabled injection of fresh CR electrons from the thermal pool
for these simulations in order to avoid contamination
by CRs that would have been injected in ICM shocks. 

\subsection{Synthetic Observations}
\label{synthobs}
Synthetic observations of the simulations were calculated following techniques  similar to those described in \citet{mend11a}.  We focus on
two models for optically thin emission; thermal bremsstrahlung from the hot plasma and synchrotron radiation from the CR electrons.  X-ray emission in an energy band typified by
\emph{Chandra} and low frequency radio emission similar to those obtainable by \emph{LOFAR} are of particular interest. We
note that, although emission due to inverse Compton scattering of CMB photons is simple to include, it is negligible for
the energy/frequency range of interest with the CR electron numbers we assume here.

The thermal bremsstrahlung emissivity was calculated in every computational zone as
\begin{align}
j_{\nu_{local}} = 5.4\times10^{-39}\,g_{ff}(\nu_{local},T_{e}) \times \\ \nonumber Z_{i}^{2}\frac{n_{e}n_{i}}{T_{e}^{1/2}}e^{-h\nu_{local}/T_{e}}
\text{ erg cm}^{-3}\text{ s}^{-1}\text{ sr}^{-1}, \label{f:brem2}
\end{align}
where $\nu_{local} = \nu_{obs}\left(1 + z\right)$ for a redshift of $z$, $T_{e}$ is in keV and all other quantities are in cgs units. 
The free-free Gaunt factor, 
$g_{ff}$, was computed by interpolation from the values calculated for plasma with typical ICM properties in Table 1 of \citet{nozawa}.  
For simplicity,
we assumed a fully ionized, $Z_{i} = 1$ pure hydrogen gas with an ideal equation of state.  The pure hydrogen assumption, however, does not match the 
abundances that
were used to define the composition of the baryons described in \S \ref{cluster_setup}.  Although formally the Gaunt factor does depend on the ion charge 
$Z_{i}$, the modifications to $g_{ff}$ for typical temperatures and observations energies of these simulations for a $Z_{i} = 2$ gas were only a 
few percent from Table 
1 of \citet{nozawa}.  The remaining dependence is the $Z_{i}^{2}$ term in Equation \ref{f:brem2}.  Since the abundance is uniform on the grid, this 
only 
sets a normalization for bremsstrahlung emission.  Finally, only for the thermal emission computation, the mean molecular weight $\mu$ = 0.5 was used to define the
temperature in a zone as $T_{e} = T_{i} = T(keV)= \mu P m_{H}/(1.602\times10^{-9}\rho)$, where $P$, $m_{H}$ and $\rho$ are all in 
cgs units.  To simulate the bandwidth of real instruments, Equation \ref{f:brem2} was numerically integrated over a range of frequencies.  Unlike the
technique used in \citet{mend11a}, where emission from zones with AGN plasma was clipped for numerical reasons, 
we do not eliminate emission from those zones in this study.  Significant and complex mixing between ICM and AGN plasma in these simulations prevented a
clear cutoff based on the passive color tracer, $C_{j}$.

Synchrotron emissivity was calculated using the local CR spectrum and was given by \cite{jos74} as
\begin{align}
\epsilon_{\nu_{local}} = j_{\alpha}\frac{4\pi e^{2}}{c}f_{p_{c}}p_{c}^{q_{c}}\left(\frac{\nu_{B}^{\alpha + 1}}{\nu_{local}^{\alpha}}\right),
\end{align}
where $f_{p_{c}}$ is the CR phase space density at momentum $p_{c}$, $q_{c}$ is the CR spectral slope at $p_{c}$ and $j_{\alpha}$ is a constant
of order unity.  The spectral index is related to the CR spectral slope as $\alpha = (q_{c} - 3)/2$.  The critical momentum is calculated 
in units of $m_{e}c$ as $p_{c} = \sqrt{2 \nu_{local} / (3 \nu_{B})}$, where the gyrofrequency is given by 
$\nu_{B} = e B / (2 \pi m_{e} c)$ and $B$ is the magnitude of the magnetic field in the plane of the sky.

A ray casting engine was used to convert volume emissivities into images with the source set at a user-defined distance.  In the engine, rays were
cast normal to an image plane that was rotated to a user-defined orientation with respect to the grid.  Trilinear interpolation was used to
determine the emissivity at regular intervals along the ray.  The sum of emissivities along a line of sight determined the intensity, and
the intensity was then converted into a flux by applying the appropriate redshift correction and multiplying it by the solid angle of a pixel determined by the luminosity distance $D_{L}$
of the source.  For both X-ray and radio observations presented here, the computational domain was set at a redshift of 
$z = 0.0594$ ($D_{L}$ = 240 Mpc), which is approximately the same as the Hydra Cluster \citep{wise07}. 


\section{Cluster Properties: Initial Conditions}
\label{cluster_cond}
\begin{figure*}
\centerline{\includegraphics[height=.7\textheight]{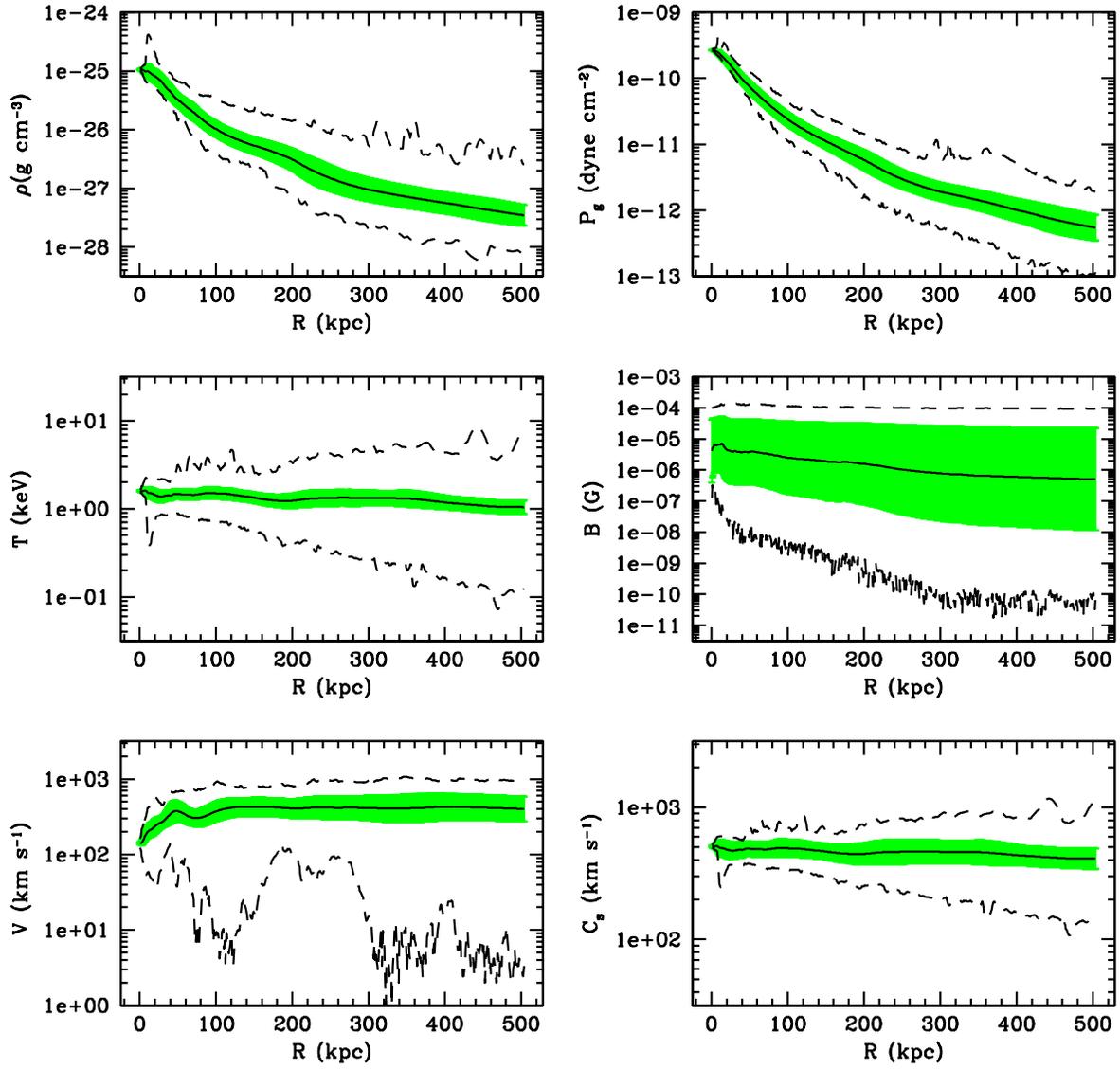}}
\caption[Fig 3 Radial profile of ICM properties from the initial conditions.]{Radial profile of ICM properties from the initial
conditions.  The mean with standard deviations are the \emph{solid} lines surrounded by the colored band,
and the minimum and maximum values at each radial sample are the \emph{dashed} lines.\\
\emph{Note}: the lower variation values for the magnetic field magnitude were consistent with zero for some radii and were then truncated.}
\label{fig:cluster_init}
\end{figure*}

\begin{figure*}
\centerline{\includegraphics[height=.55\textheight]{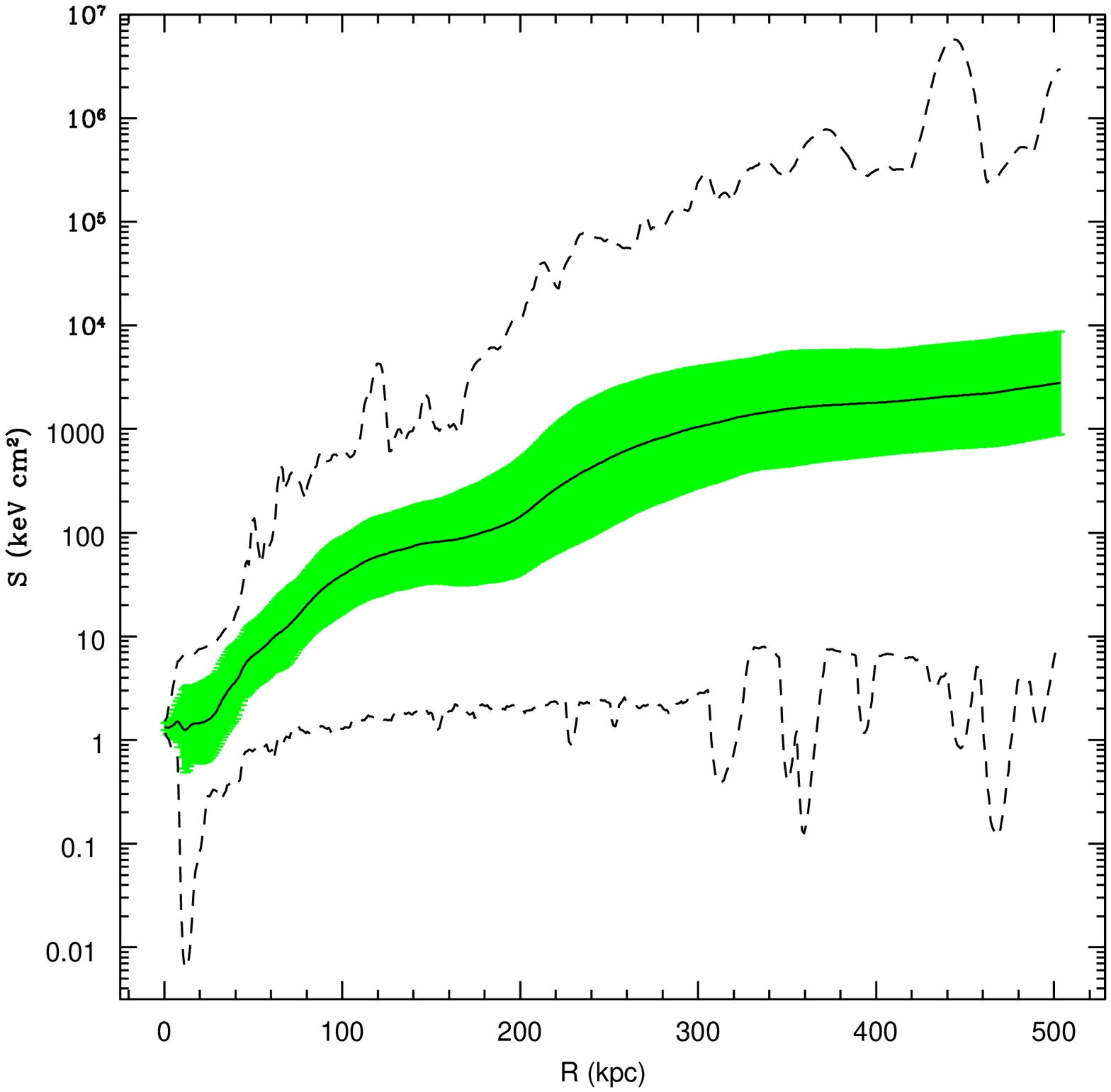}}
\caption[Fig 4 Radial profile of ICM entropy from the initial conditions.]{Radial profile of ICM entropy from the initial
conditions.  The mean with standard deviations are the \emph{solid} line and color band,
and the minimum and maximum values at each radial sample are the \emph{dashed} lines.}
\label{fig:cluster_entropy}
\end{figure*}

\begin{figure*}
\centerline{\includegraphics[height=.27\textheight]{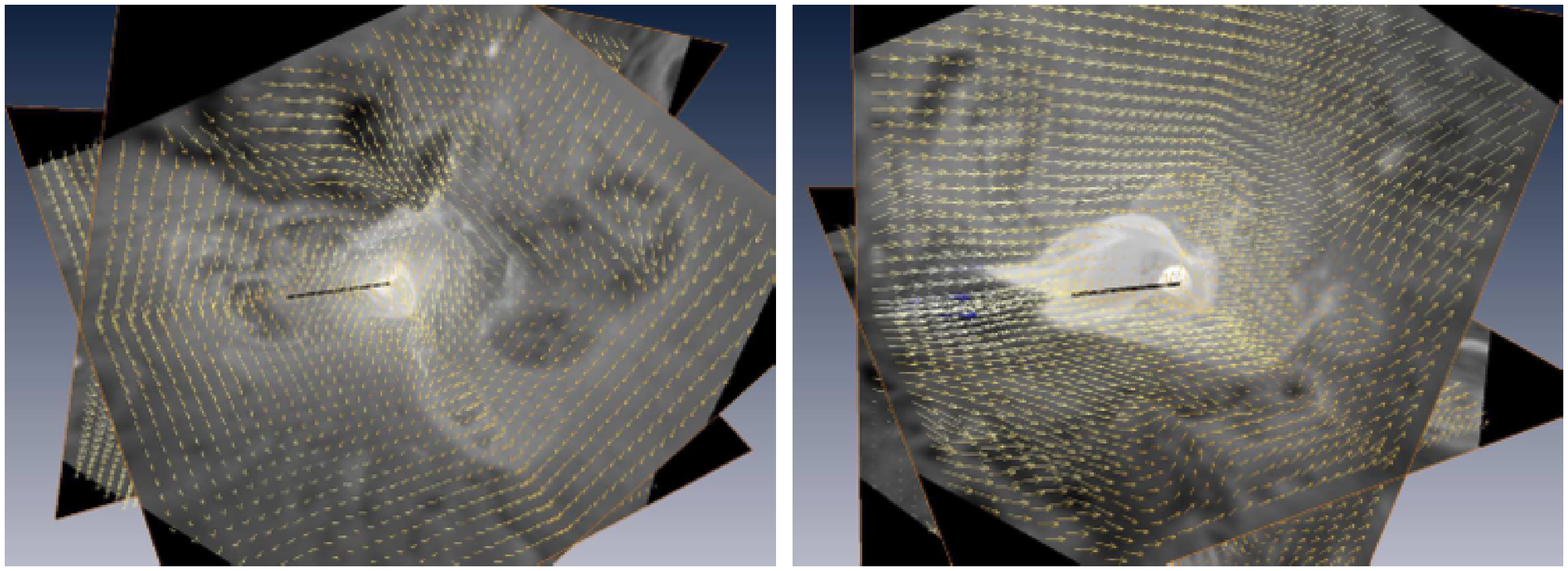}}
\caption[Fig 5 Slices of entropy with velocity vectors overlaid from the initial conditions.]{Slices of gas entropy, $S = P/\rho^{\gamma}$,
in three orthogonal planes from the initial conditions.  The panel on the \emph{left} shows the initial conditions rotated such that
the jet axis from {\bf R1}, seen as a \emph{dark black} line from cluster center, is at the intersection of two of the planes.  The panel
on the \emph{right} is the same for {\bf R2}.  Light to dark colors correspond to low to high values of gas entropy.  Overlaid in both
panels are velocity vectors in the same three planes as entropy.  Longer vectors correspond to higher speeds.}
\label{fig:cluster_entropy_slices}
\end{figure*}

To set the stage for analysis we outline now and illustrate in Figures \ref{fig:cluster_init} -  \ref{fig:cluster_entropy_slices} the initial properties of the ICM for our cluster before the jets were launched.  Figure 
\ref{fig:cluster_init} shows from the simulation initial conditions azimuthally averaged radial profiles for gas density and pressure (along with the derived temperature and sound speed),
as well as the ICM flow speed with respect to the DM core and the magnetic field strength. 
The variance and extremes of each quantity are also shown. The cluster has
a dense core with a central ICM density of $\rho_{ICM,0} = 1.05 \times 10^{-25}$ g cm$^{-3}$ and pressure 
$P_{ICM,0} = 2.69 \times 10^{-10}$ dyne cm$^{-2}$.  This gives a core temperature of $T_{ICM,0}$ = 1.6 keV, for 
the assumed mean molecular weight of $\mu$ = 0.6, and a sound speed of $c_{ICM,0} = 653$ km s$^{-1}$.
The azimuthally averaged temperature falls off slowly with radius, which is typical for non-cool core clusters \citep{hudson10}.  This is not surprising,
as radiative cooling was not included in the GADGET simulation that evolved the cluster to this time.
The average magnetic field magnitude in the core is $B_{ICM,0} = $ 4.2 $\mu G$, which corresponds to a relatively high plasma 
$\beta_{ICM,0} = 8\pi P_{ICM,0} / B_{ICM,0}^{2} = 382$.  
Figure \ref{fig:cluster_entropy} shows a radial profile of the ICM entropy for the initial conditions.  This cluster did have
a low entropy core, which is a standard property of SPH simulations like the one that produced the cluster \citep[\emph{e.g.,}][]{mitchell09}.

Although this cluster was selected especially because its last major merger was  several
gigayears in the past and at first appearance it seems relatively relaxed,
closer inspection reveals significant  ICM ``weather'' characteristic of
sloshing behaviors, thus, reminding us that g676 has continued to interact with
smaller DM halos and to accumulate mass slowly.
Note first from Figure \ref{fig:cluster_init} that
the mean magnitude of the ICM velocity in the inner core is 
$v_{ICM,0} = 142$ km s$^{-1}$ (Mach 0.2), and that it increases to $\approx$ 400 km s$^{-1}$ by 100 kpc from the cluster center.  
For a typical sound speed of 450 km s$^{-1}$ outside 100 kpc, these motions
approach Mach 1.  They
are primarily relatively large scale flows across the cluster, since by contrast, 
\cite{zhura11} report isotropic, random velocities with respect to the 
mean of only $\sim 50~ \rm{km~s^{-1}}$ in the core of this cluster,
increasing to around  $\sim 100~ \rm{km~s^{-1}}$ outside a 100 kpc radius. The turbulent pressure in the core
region ($P_{turb}\sim 3\times 10^{-12}$ dyne cm$^{-2}$), therefore, is less than 1\% of the core thermal pressure (Figure \ref{fig:cluster_init} shows profiles of density and thermal pressure).

Further indications of the sloshing behavior are found in Figure \ref{fig:cluster_entropy_slices}, which shows orthogonal slices of gas entropy, $S = P/\rho^{\gamma}$.
The initial cluster conditions were rotated such that the jet axis from {\bf R1} (see \S \ref{sim_results}) is in two of the planes on the left panel, and the panel on the
right was created in the same fashion for {\bf R2} (see \S \ref{sim_results}).  An inverse colormap was used, where lighter colors correspond to lower entropy.  The jet axis is shown as a 
dark line from the cluster center.  Overlaid are ICM velocity vectors whose length corresponds to the gas speed.  Both views show comples flow patterns and
highly asymmetric entropy distributions, which are classic signatures of sloshing \citep{zuhone10}.

\begin{figure}
\centerline{\includegraphics[height=.3\textheight]{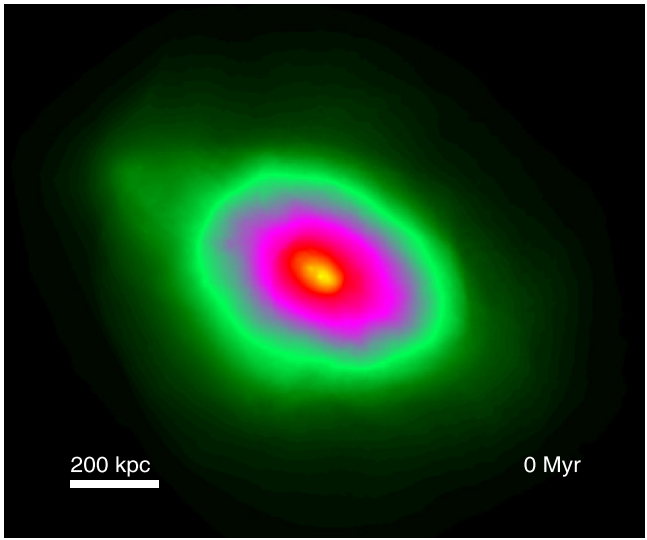}}
\caption[Fig 6 X-ray observation from 0.5 to 8 keV of inital conditions.]{Synthetic X-ray observation from 0.5 to 8 keV of the initial cluster
conditions.  The image is scaled logarithmically over $\approx$ 3 orders of magnitude. The line of sight for this observation was along the jet axis
from the {\bf R1} simulation described in \S \ref{run1_results}.}
\label{fig:cluster_xray}
\end{figure}

Despite these flow patterns this cluster does still qualify by common observational measures as an example of a relaxed system.
A synthetic X-ray observation of the initial cluster in the 0.5-8 keV band is shown in Figure \ref{fig:cluster_xray} along a line of sight roughly
orthogonal to the mean ICM velocity in the cluster core.  This line of sight would also be along the jet axis from the {\bf R1} simulation (see \S \ref{sim_results}).  Although there is
some asymmetry of the cluster core for this particular line of sight, other observational orientations lead to
more symmetric images. As a quantitative measure we construct from this image the concentration parameter, $c$,  sometimes used to
separate relaxed clusters from those that have recently undergone mergers  \citep[\emph{e.g.,}][]{santos08};
\begin{align}
c = \frac{S(R < 100 kpc)}{S(R < 500 kpc)},
\end{align}
where $S$ is the X-ray flux in the aperture.  A relaxed cluster would have a more concentrated core, so a higher value of $c$,
 than a cluster that was recently disturbed in a merger.  \citet{cassano10} find that a value $c > 0.2$ separates clusters without 
radio halos (not recently disturbed) from those with radio halos and a recent major
dynamical disturbance.  From the X-ray observation in 
Figure \ref{fig:cluster_xray},
the initial cluster conditions used for this study have a concentration parameter of $c = 0.66$, so the X-ray profile would not be considered dynamically disturbed 
by this observational method.  This measurement was repeated for two other orthogonal lines of sight through the cluster with variation in $c$
of less than 1\%.
For reference we note that it has been pointed out \citep{cassano10} that the concentration parameter is less sensitive to projection effects than the so-called power 
ratio based on multipole expansions of the X-ray brightness distribution \citep[\emph{e.g.,}][]{buote95,jeltema05,ventimiglia08,bohringer10,cassano10}.

\clearpage


\section{Jet Simulation Results}
\label{sim_results}
Two AGN outflow simulations for the parameters defined above in \S \ref{jet_setup} were run with the jet cylinder at different orientations on the grid.
Both AGNs were place at rest in the cluster center in order to distinguish the influences of the dynamic ICM from host galaxy motions within the cluster.
The first simulation, {\bf R1}, had the jet cylinder orientation chosen arbitrarily, but so that the jet axis was not along a
primary grid axis. As it turned out the jets were directed roughly orthogonal to the mean ICM velocity inside a 200-300 kpc cluster radius, which
represents the final scale of the cavities formed in the ICM by the AGN outflows.
The second simulation, {\bf R2}, had the jet cylinder in the orthogonal plane to {\bf R1} and in a direction more nearly aligned with the mean interior ICM flows.
The total elapsed time for each simulation was 200 Myr.  The jets cycled on and off with a 50\%
duty cycle for 6 periods of 26.2
Myr per cycle and then left off after 160.5 Myr of this activity, allowing the AGN inflated bubbles to evolve without more energy injection for about an additional 40 Myr.
 We note in passing that a steady jet of the same properties over this same
time interval would propagate off the grid. A steady jet would create a structure roughly twice as long as those produced in the two simulations we present.  
Additionally, the X-ray cavity produced by a steady jet would be significantly less round in shape as compared to those produced by the intermittent jets
in the simulations shown here. 

\subsection{Simulation 1 ({\bf R1}) Evolutionary Summary}
\label{run1_results}
\begin{figure*}
\centerline{\includegraphics[height=.82\textheight]{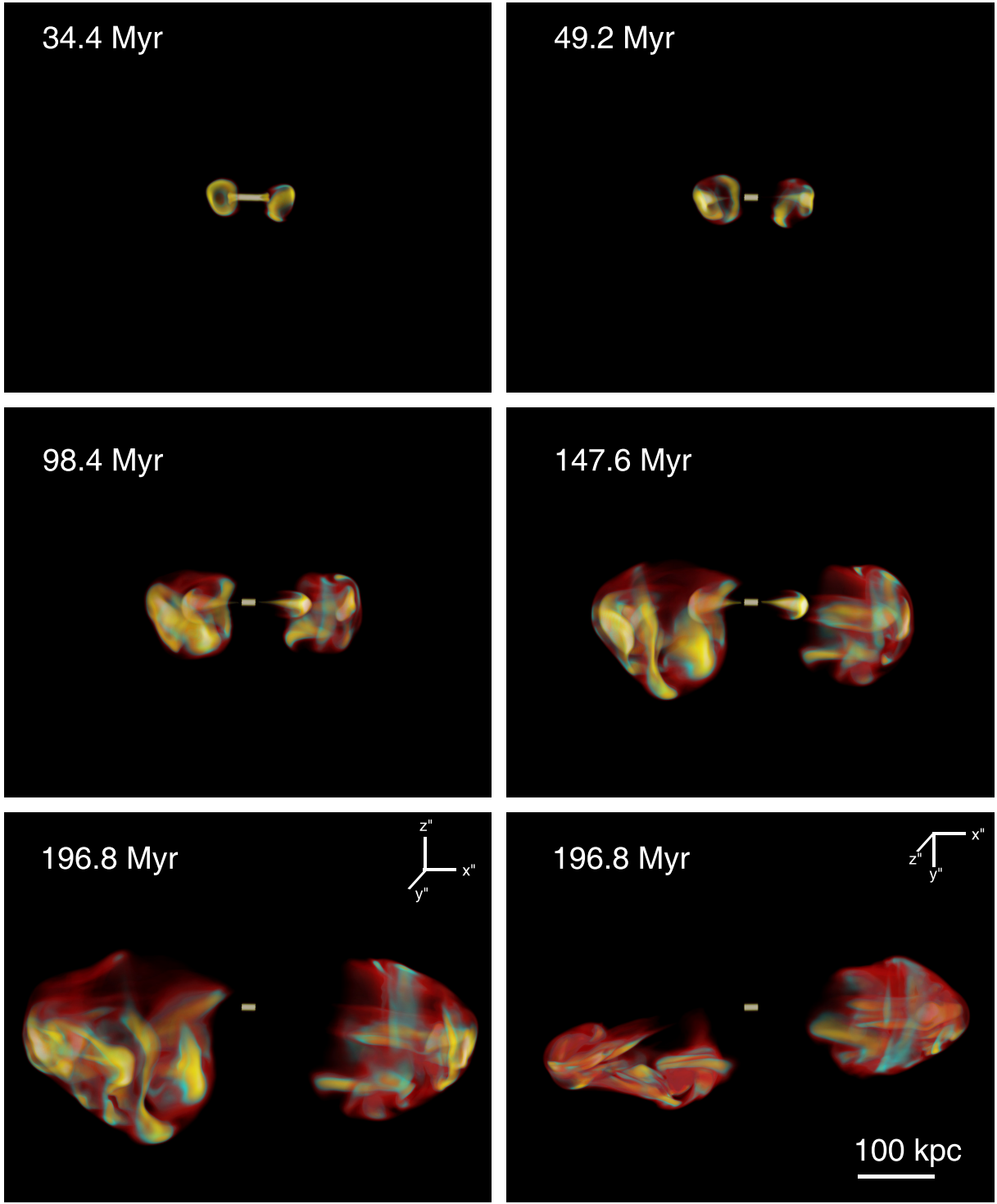}}
\caption[Fig 8 Color volume rendering from {\bf R1}.]{Volume rendering of the color variable, $C_{j}$, from {\bf R1}.  For color the color version of
this figure, the color map has ICM plasma
in \emph{red}, equally mixed ICM and AGN plasma in \emph{turquoise}, and predominantly AGN plasma in \emph{yellow} to \emph{white}.
The grayscale version of this figure shows AGN plasma in \emph{white} and ICM plasma in \emph{black}.}
\label{fig:run1_color}
\end{figure*}

Figure \ref{fig:run1_color} shows color, $C_{j}$  (= jet mass fraction), volume renderings from the {\bf R1} simulation at five representative times (t = 34.4, 49.2, 98.4, 147.6 and 196.8 Myr). 
The views are orthogonal to the jet axis. All
the panels project the same view except the lower right, as indicated by the stick diagrams in the two bottom panels.
A line segment in the lower right panel indicates a projected 100 kpc length. To aid structural registration the location of the jet cylinder is shown
 independent of jet activity. 

At 34.4 Myr (upper left), the AGN is active for the
second time in the simulation, so the jets are collimated, and there are small lobes formed from the first active AGN phase.  
As mentioned, these jets drove through a strong cross flow in the ICM associated
with sloshing motions. Already at this time
the lobes are obviously asymmetric and no longer aligned with the jet axis.  At 49.2 Myr the AGN is inactive, and it is apparent that the AGN
plasma is concentrated at the ends of the lobes.  The panels at 98.4 Myr and 147.6 Myr both show times when the jets are on.  In both cases
the visible jets clearly are not as well collimated as they were at 34.4 Myr.  The ICM pressure near the launching region is approximately a factor of two lower
at these times than the value at 34.4 Myr.  Recall from \S \ref{jet_setup} that the jets were launched with a fixed pressure related to the initial ICM, so the jets are 
now over-pressured as they emerge, causing them to expand. Since the internal jet Mach number is low ($M_j \approx 1.2$), 
re-collimation is slow. In addition,  bow shocks have formed ahead of the new outflows, as well as jet termination shocks, further enhancing
the dispersion of jet plasma into the cavity associated with the previous activity. The asymmetry between the lobes and their misalignment
with the jet outflows  are quite obvious, especially in the final views at 196.8 My, sr.
The lobes extend roughly the same distance 
from the cluster center (300 kpc), but there are different offsets of the lobes from the jet axis.  The most significant differences between the
two lobes are the deflection in roughly the $-\hat z^{\prime\prime}$ direction, so that viewed roughly along the $\hat y^{\prime\prime}$ axis they present a ``C-shaped'' morphology, 
and from another, orthogonal perspective, a flattened shape of the left lobe compared to the right lobe, as seen in the bottom right panel of Figure 
\ref{fig:run1_color}.  Finally, we note that the AGN plasma is very nonuniformally distributed inside the lobes, with some significant mixing of ICM and AGN plasma 
in several regions, made evident by the gray (turquoise for color figures) regions in all the figure panels.
Such mixing should have various observational consequences, including the Faraday
rotation behaviors along sight lines through the lobes, as we will discuss
in a subsequent publication.

\begin{figure*}
\centerline{\includegraphics[height=.33\textheight]{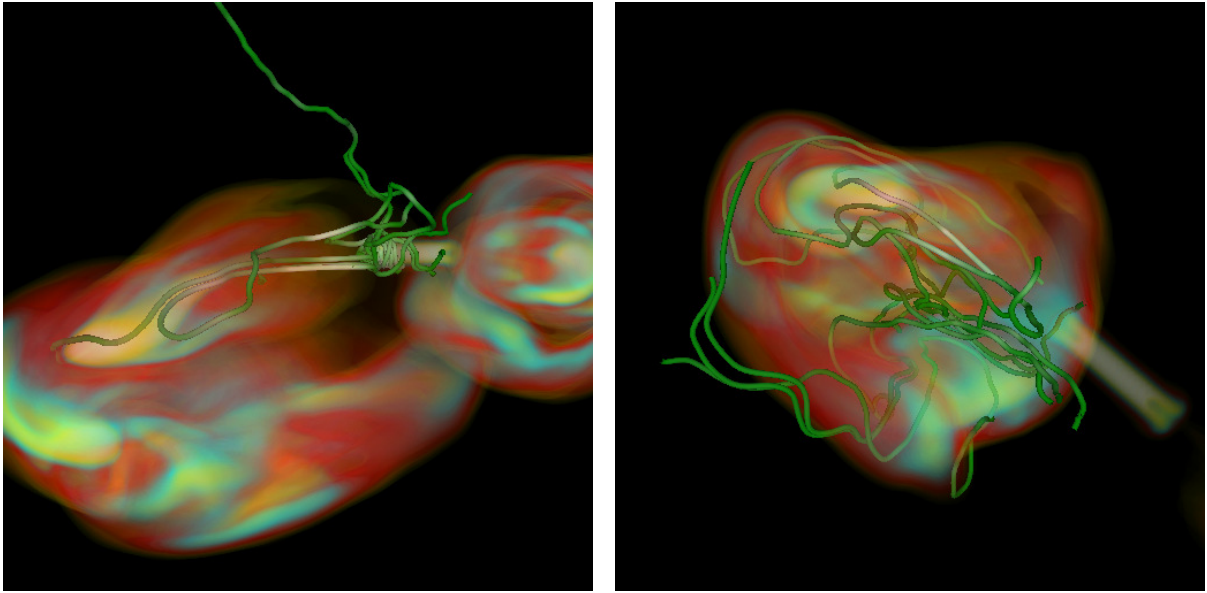}}
\caption[Fig 9 Magnetic field lines in {\bf R1}.]{Volume rendering of passive tracer $C_{j}$  and selected magnetic field lines from {\bf R1} during the active AGN phase at 137.8 Myr. Field lines are colored by magnitude with brighter field line colors representing stronger fields.  \emph{left}: A view showing the tightly coiled jet field near the launching region.
ICM field lines are seen stretched over the AGN plasma from a previous AGN cycle. \emph{right}: Field lines in this view within the leftward jet lobe in
the other panel are seen following the outline of the lobe.}
\label{fig:run1_bfield}
\end{figure*}

To that point, however, we mention here that magnetic fields, while not dynamically dominant in these simulations, are
essential in the production and transfer of radio emissions. In those roles they can help to illuminate dynamical properties of
the AGN/ICM interactions.  Figure \ref{fig:run1_bfield} shows volume renderings from the {\bf R1} simulation of the color, $C_{j}$, along with selected magnetic 
field lines colored dark to light by magnitude during an active AGN phase at 137.8 Myr.
In the left panel the jet to the right is coming  towards the observer and the left jet is directed away.  
From this perspective the AGN outflows are deflected away from the observer.  Also in the left panel, ICM
field lines can be seen stretched and dragged outward with AGN plasma into the leftward lobe formed from previous activity.  Those 
 ICM field are actually draped around the AGN plasma. 
The field lines colored white show where the stretching from the jet outflow has most intensified the field.  OJ10 reported a similar result for their intermittent
jet simulations.  There is a stray magnetic field line seen in the left panel of Figure \ref{fig:run1_bfield}.  Despite attempting to isolate field within the jet,
this line is rendered due to the blending of the jet and ICM magnetic fields in the collar region around the jet cylinder.
In the right panel the magnetic field inside
this  same lobe is shown from a slightly different perspective wrapping around regions consisting primarily of AGN plasma.  Field lines are also evident tracing the outline of the lobe.

\subsection{Simulation 2 ({\bf R2}) Evolutionary Summary}
\label{run2_results}
\begin{figure*}
\centerline{\includegraphics[height=.82\textheight]{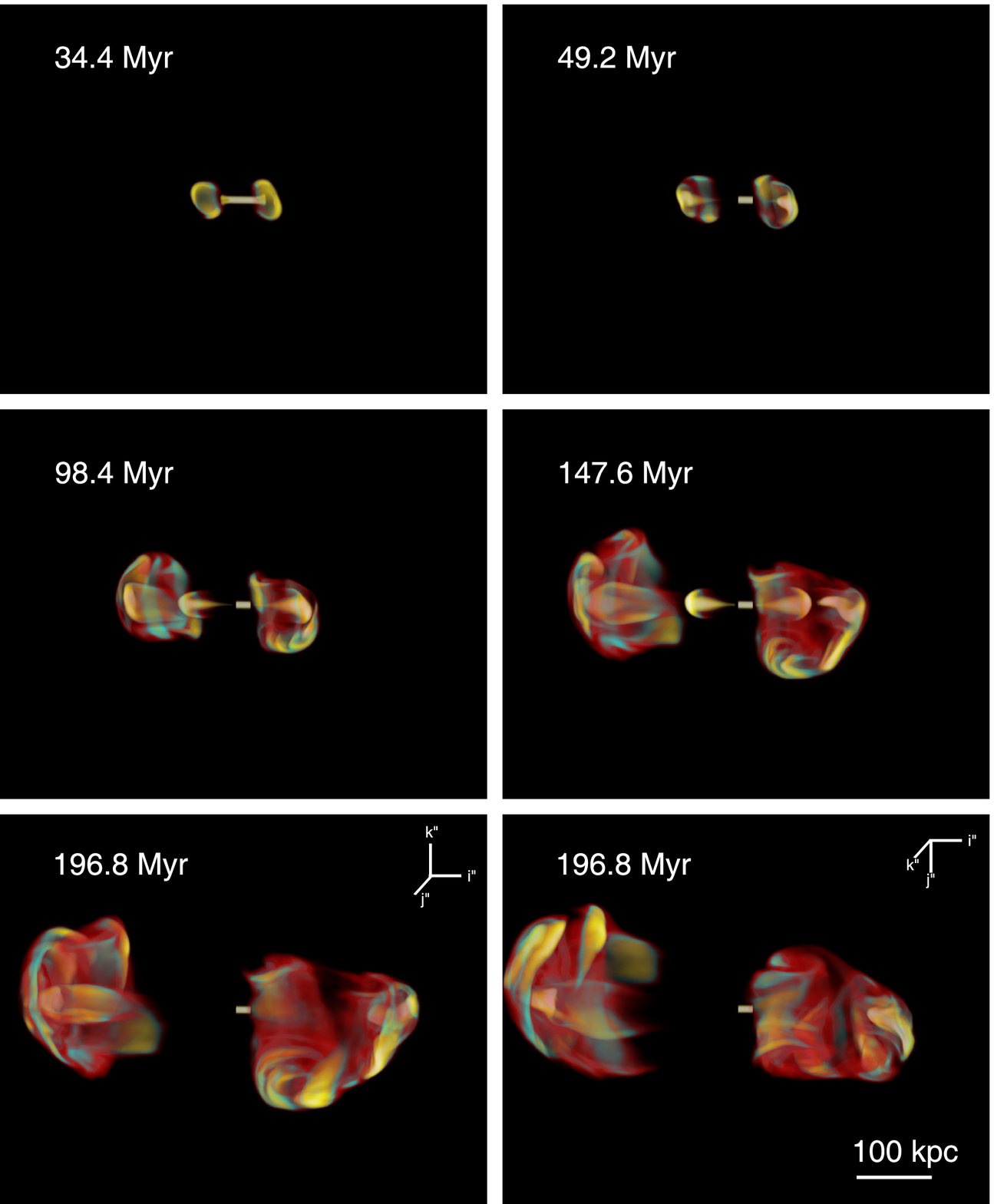}}
\caption[Fig 10 Color volume rendering from {\bf R2}.]{Same as Figure \ref{fig:run1_color} for {\bf R2}. }
\label{fig:run2_color}
\end{figure*}

The evolution of the {\bf R2} simulation is shown in Figure \ref{fig:run2_color},
again through volume renderings of the color $C_{j}$. 
Many of the same morphological characteristics of {\bf R1} visible in 
Figure \ref{fig:run1_color} can be seen in this rendering of {\bf R2}, as well.  Just as in the {\bf R1}
case the {\bf R2} jets appear well
collimated early in the simulation, but again they over-expand as the  propagate
through the lower pressure cavities.  
The
AGN plasma  in the lobes has a qualitatively similar distribution to {\bf R1}. 
Again the AGN plasma is primarily deposited near the outer edges of the lobes.  
The most striking difference between
the {\bf R1} and {\bf R2} morphologies is that while the lobes in the {\bf R1} simulation extended about the same distance from the AGN, the lengths of the two lobes in {\bf R2} are of very different length.  The lobe on the right side of Figure \ref{fig:run2_color} does not extend nearly as far into the cluster as the lobe on the left side in these images. This is evident even in the early, 34.4 Myr image. The asymmetry continues to increase throughout the
simulation. At 196.8 Myr the difference in extent is approximately 100 kpc
or roughly 50\%. Recall that these jets are directed in an orthogonal direction to those
in the {\bf R1} simulation. In fact, the jet axis is roughly aligned parallel and anti-parallel to the prevailing ICM velocity in the inner cluster mentioned previously. The relationship
between cavity morphology and cluster weather will be discussed in more detail in \S \ref{weather}.

\begin{figure*}
\centerline{\includegraphics[height=.33\textheight]{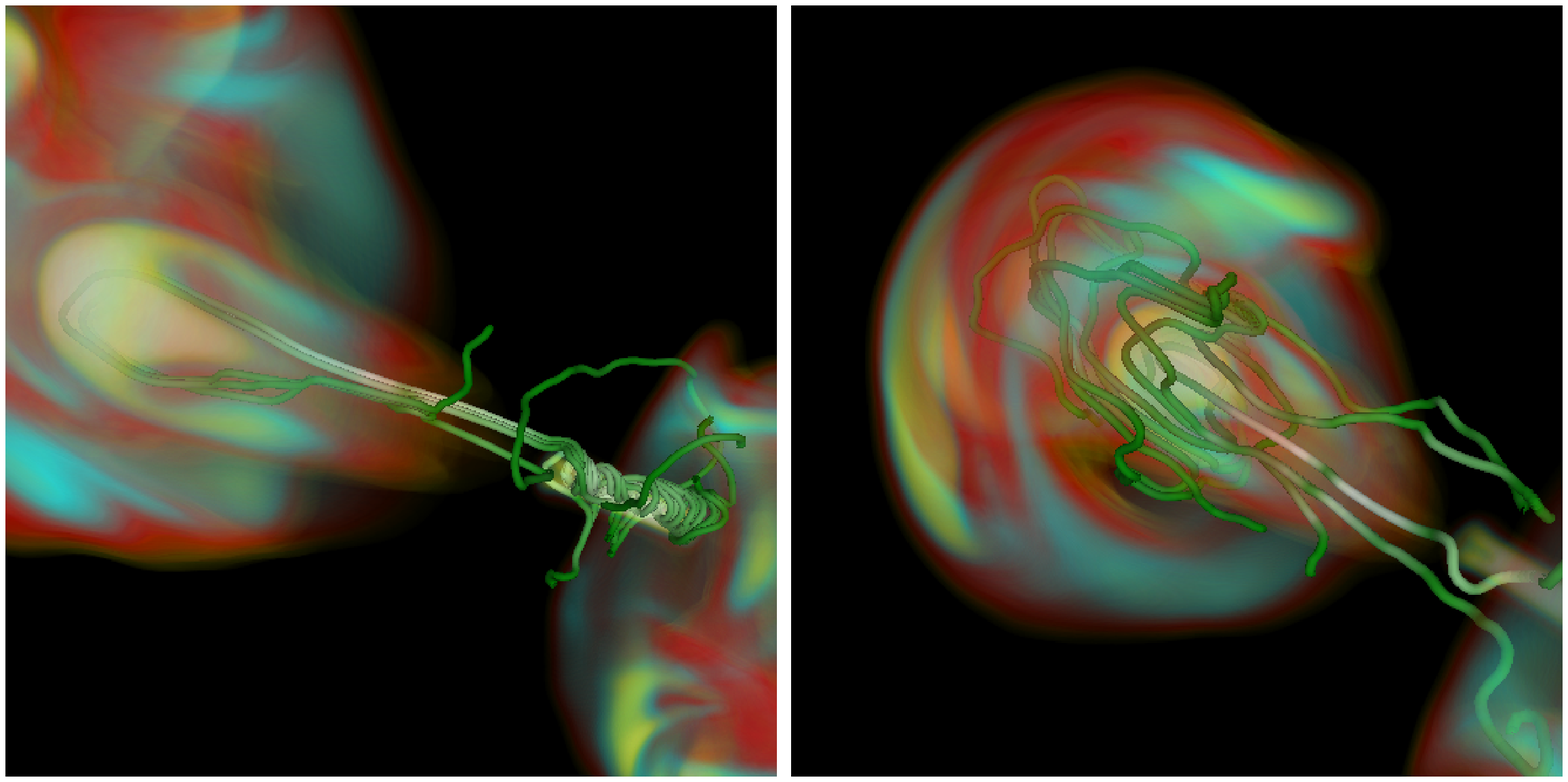}}
\caption[Fig 11 Magnetic field lines in {\bf R2}.]{Similar to Figure \ref{fig:run1_bfield} for {\bf R2}.}
\label{fig:run2_bfield}
\end{figure*}

The structure of several magnetic field lines during jet activity at 137.8 Myr from {\bf R2} are seen in Figure \ref{fig:run2_bfield}.  Similar to the 
structures in Figure \ref{fig:run1_bfield}, the toroidal jet field is shown in the left panel, along with ICM field lines that have been stretched and wrapped
over AGN plasma from the last period of activity.  Magnetic field lines inside the leftward lobe are visible in the right panel. Those lines also follow the
outline of lobe structures, as they did in the {\bf R1} simulation.  The fields seen here appear to be less tangled than those seen in Figure \ref{fig:run1_bfield}.  This same lobe is the one on the left-hand side of the images in Figure \ref{fig:run2_color}. Its distortions are not as extreme as the deflected jet
and lobe shown from {\bf R1} in Figure \ref{fig:run1_bfield}.


\subsection{Synthetic X-ray Observations}
\label{dyn_xray_obs}
\begin{figure*}
\centerline{\includegraphics[height=.72\textheight]{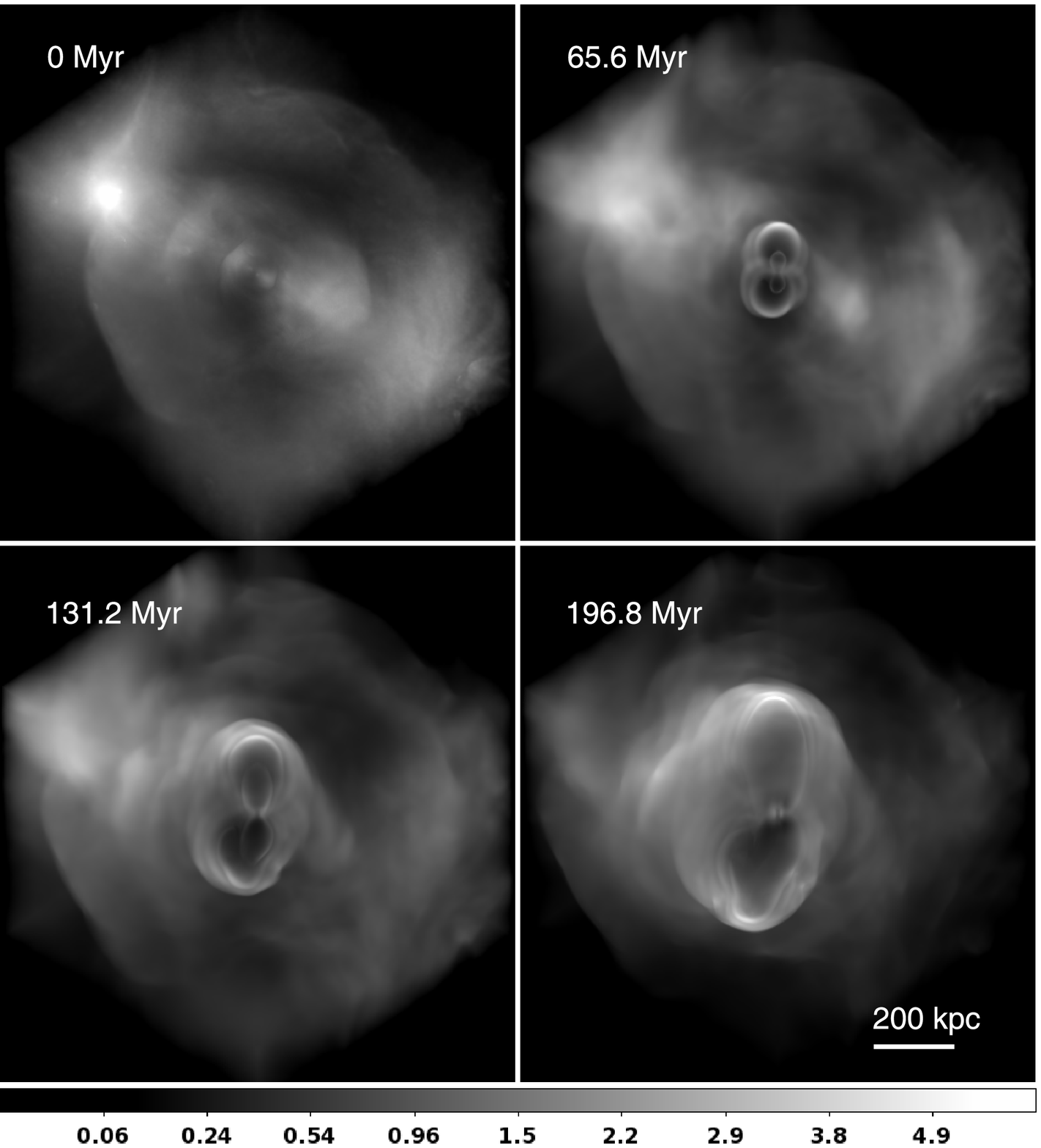}}
\caption[Fig 12 X-ray observations from 0.5 to 8 keV of {\bf R1} divided by $\beta$-profiles.]{Synthetic X-ray observations from 0.5 to 8 keV from four epochs
of {\bf R1}.  Each observation was divided by a best-fit $\beta$-profile to emphasize the X-ray cavities.}
\label{fig:run1_xray}
\end{figure*}

Figure \ref{fig:run1_xray} shows 0.5-8 keV X-ray observations of {\bf R1} at various epochs taken along a line
of sight that was perpendicular to the jet axis.  The resolution is 1 arc sec per pixel at the assumed 240 Mpc distance, and each observation was divided by the best-fit 
double $\beta$-profile \citep[see][]{mend11a} to emphasize the X-ray
cavities.  In these images, the lower X-ray cavity is associated with the \emph{left-hand} lobe in Figure 
\ref{fig:run1_color}.  The observation of the initial conditions  shows significant departures in surface brightness from the double $\beta$-profile, 
particularity
at large distances from the cluster center 
(see, also Figures \ref{fig:cluster_entropy_slices} and \ref{fig:cluster_xray}) reminding us, 
once again that the ICM is not truly relaxed and the cluster is not isolated.  
At 65.6 Myr, after two full AGN
periods, the jets are active and and the inner pair and outer pair of cavities are seen separated by a thin bright rim.  These features are similar
to the multi-cavity system in Abell 2052 \citep{blanton09}.  The observation at 131.2 Myr shows the same morphology with the pairs of cavities
on larger scales.  Five complete AGN cycles have been executed by this time.  Ripples are seen emanating from the outer pair of cavities, which
are outlined by bright rims that are $\sim$ 30\% hotter than the surrounding gas.  The ripples were seen for the intermittent simulation presented 
in \citet{mend11a}, and they are similar
to the ripples observed in the Perseus Cluster \citep{fabian03}.  Additionally, bright rims hotter than their surroundings have been observed in
Centaurus A \citep{kraft03, kraft07} and NGC 3801 \citep{croston07}, but for other objects, the bright rims are, in fact, cooler than their 
surroundings \citep{diehl08}.

The outer {\bf R1} cavities at 131.2 Myr are distorted and no longer aligned with
the axis of the inner pair of cavities.  These distortions are largely due to cluster ``weather'', as mentioned above and discussed more fully in \S \ref{weather}. The approximately Mach 2 bow shock is well separated from the cavities at this time.  By the time of
the final observation, at 196.8 Myr, the AGN has been inactive 
for 40 Myr, and there
is no longer an apparent pair of inner cavities.  The cavity system appears to  include only one pair that extend $\sim$ 300 kpc from the cluster center
and are $\sim$ 150 kpc wide.  The cavities are still outlined by bright rims, and the ripples can still be seen.  The lower cavity shows the
most distinct distortions.  The bow shock, now  Mach $\sim$1.3, is still visible.

\begin{figure*}
\centerline{\includegraphics[height=.72\textheight]{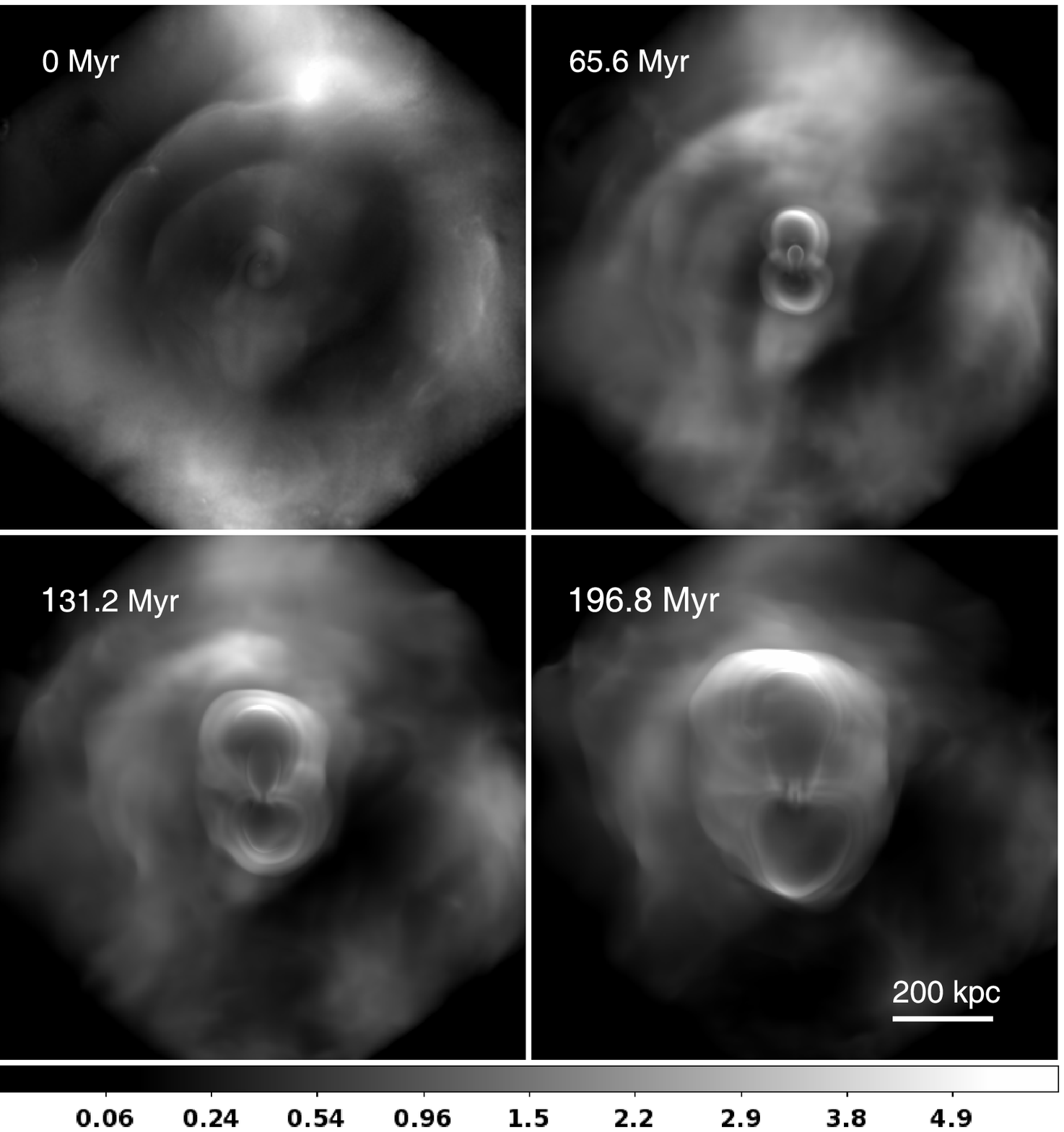}}
\caption[Fig 13 X-ray observations from 0.5 to 8 keV of {\bf R2} divided by $\beta$-profiles.]{Same as Figure \ref{fig:run1_xray} for {\bf R2}.}
\label{fig:run2_xray} 
\end{figure*}

Figure \ref{fig:run2_xray} shows 0.5-8 keV observations of the {\bf R2} simulation,  also divided by best-fit double 
$\beta$-profiles.  
In these images the 
line of sight is again perpendicular to the jet axis.  
Many of the same feature seen in the {\bf R1} observations are present, such as the ripples from
the intermittent AGN activity, bright rims around the large cavities and a bow shock with a similar Mach number to the {\bf R1} case.  A unique and very interesting
feature of the {\bf R2} observations, however, 
is that the 65.6 Myr and 131.2 Myr observations show an incomplete pair of inner cavities whereas {\bf R1} showed a complete pair.  The rim around the 
upper-inner cavity is visible in these two images, but the lower-inner cavity is minimally outlined
with a faint rim.  It is unlikely that this cavity would be seen in a real observation of such an object with finite sensitivity, and
only three of the four cavities might be identified.  In the last image at 196.8 Myr the inner cavities are absent, and only a single pair
of cavities with bright rims, that are less noticeable than those from {\bf R1}, are visible.  The lower cavity is significantly shorter and more
circular in appearance than the upper cavity. 
As  mentioned previously and discussed in more detail in \S \ref{weather}, 
this asymmetry also results from large scale ICM flows (weather) present in the 
cluster initial conditions. As mentioned previously the {\bf R2} jets are more or less aligned with this flow, whereas the {\bf R1} jets were roughly orthogonal to the 
mean interior ICM flow.

\subsection{Synthetic Radio Observations}

\begin{figure*}
\centerline{\includegraphics[height=.72\textheight]{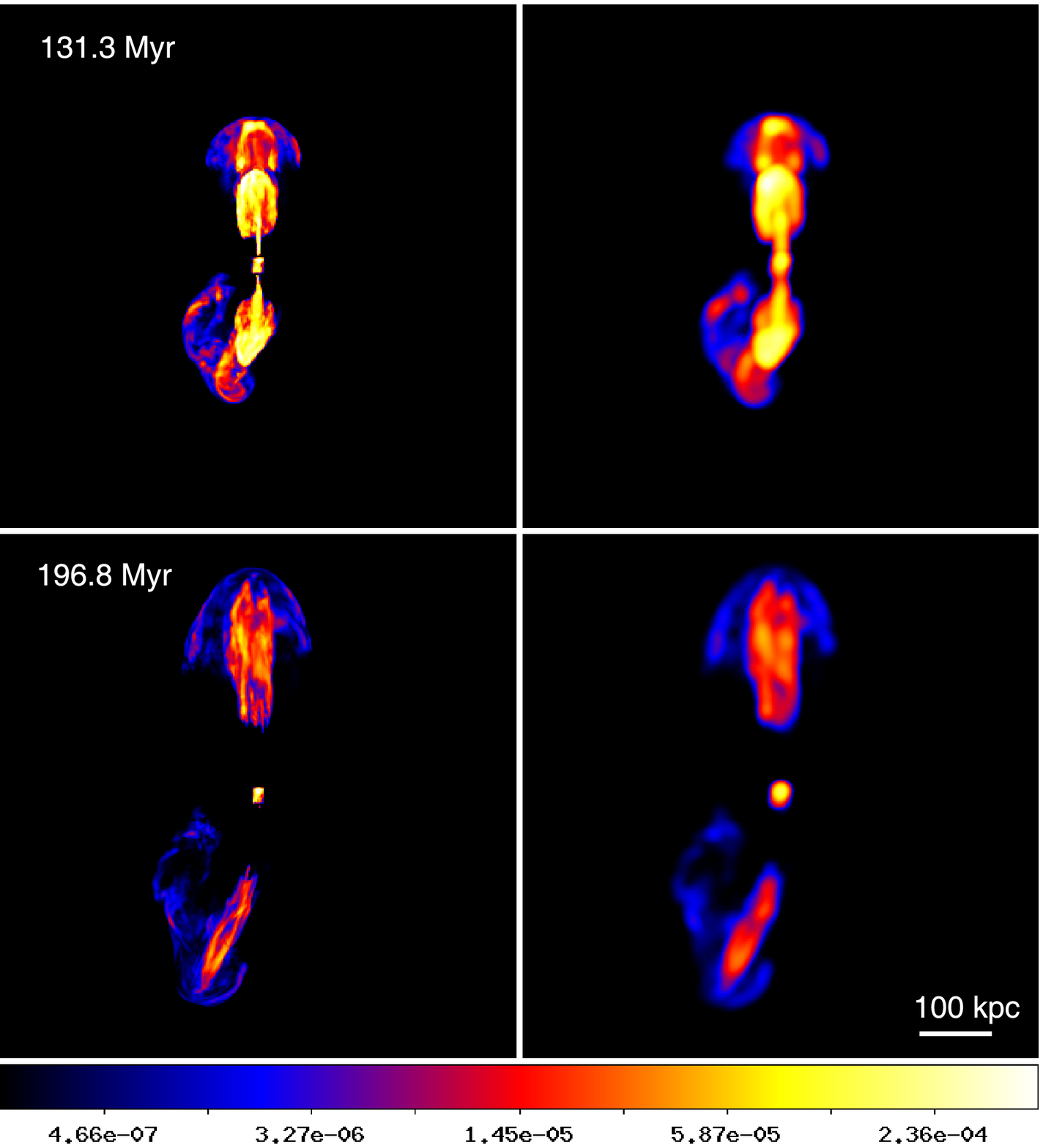}}
\caption[Fig 14 178MHz observations of {\bf R2}.]{Synthetic observations of synchrotron emission from {\bf R1} at 178 MHz.  The units on the colorbar are Jy beam$^{-1}$
with a 1.7 arcsec beam (pixel).  \emph{left}: Full resolution observation. \emph{right}: Same observation smoothed with a Gaussian kernel over
6 beams (pixels).}
\label{fig:run1_radio}
\end{figure*}

\begin{figure*}
\centerline{\includegraphics[height=.72\textheight]{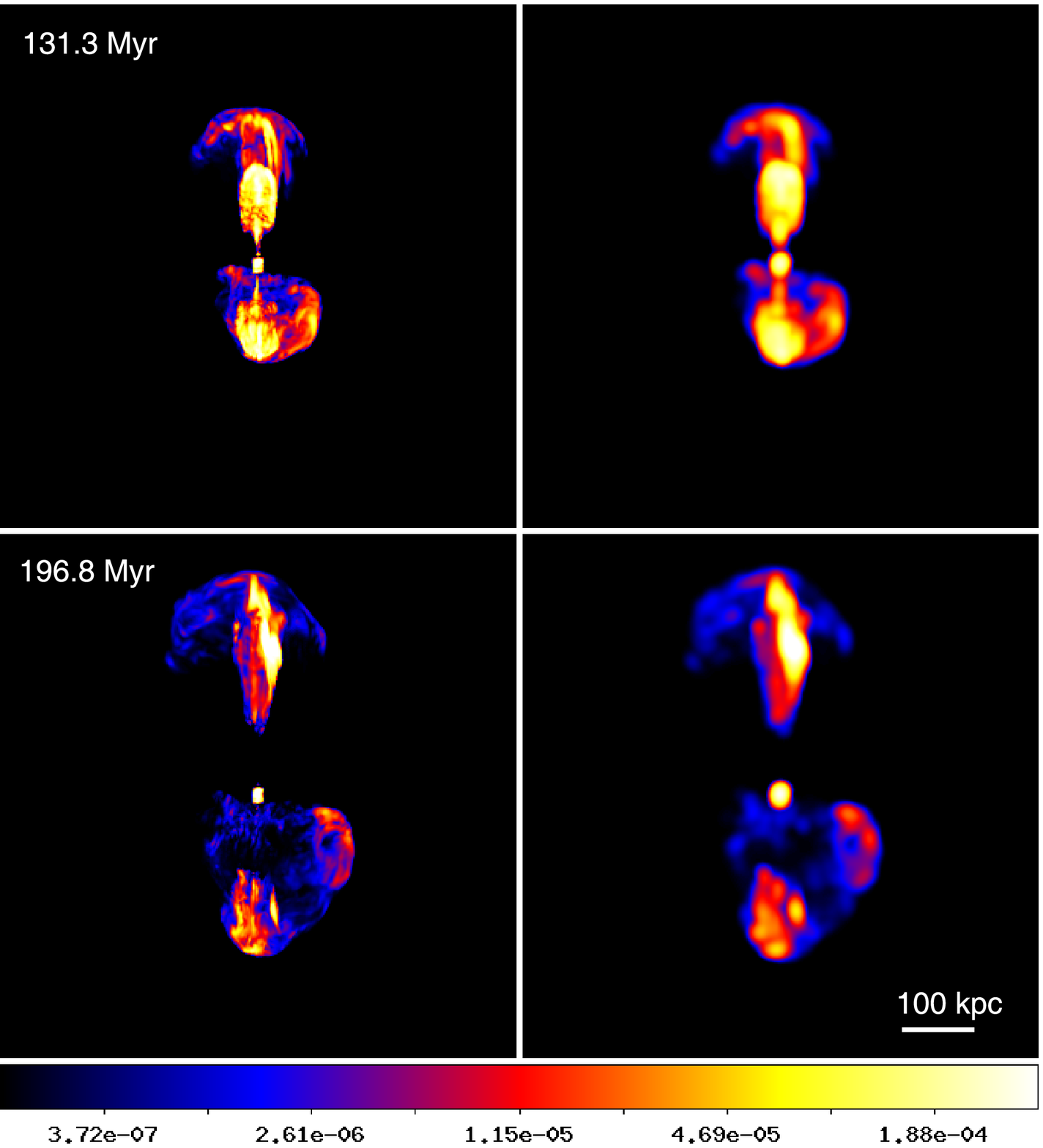}}
\caption[Fig 15 178MHz observations of {\bf R2}.]{Same as Figure \ref{fig:run1_radio} for {\bf R2}.}
\label{fig:run2_radio}
\end{figure*}

Figures \ref{fig:run1_radio} and \ref{fig:run2_radio} show at 131.3 Myr and 198.2 Myr the 178 MHz synthetic synchrotron intensity observations of the
{\bf R1} and {\bf R2} simulations, respectively.  These observations are of the same line of sight as the respective X-ray observations discussed above, but they
are of from a smaller volume to show more detail.  We note to avoid confusion that in all the images the jet launch cylinder remains artificially illuminated.
The left-hand panels in both figures were configured
with 1.7 arc sec beams (pixels) at the assumed 240 Mpc distance, and the right-hand panels are the same observations but Gaussian smoothed over 6 pixels to simulate a 10 arc sec
beam.  The line of sight through the
grid in Figure \ref{fig:run1_radio} (Figure \ref{fig:run2_radio}) is  the same as the X-ray image in Figure \ref{fig:run1_xray} (Figure \ref{fig:run2_xray}).
Such low frequency radio observations are of particular interest, because they highlight low energy CRs that provide a reasonable
tracer for AGN plasma, since their losses are not large.  For a typical field value  in the lobes of $\sim$ 1.5 $\mu$G much of the
emission at 178 MHz is from $\sim$ 3 GeV electrons, whose
radiative lifetimes against inverse Compton and synchrotron losses in
this context are  $\sim$ 200 Myr, assuming a small cluster redshift.

For both simulations at 131.3 Myr, the bright active jets are seen in the figures
penetrating their associated lobes.  Emission from
previous AGN activity extends further out, but is considerably dimmed because of radiative aging of the CR population.  
These images closely resemble a 
``double-double'' radio galaxy (DDRG) \citep{schoenmakers00} such as 3C 293, which also shows
a misalignment between the inner and outer doubles \citep{joshi11}.  At the end of the simulations, most of the DDRG resemblance is gone because
roughly 50 Myr has passed since the AGN was last active.  {\bf R1} shows a distinctly bent morphology, similar to a WAT radio source.  
The {\bf R2} observation in Figure \ref{fig:run2_radio} has a second pocket of emission in the lower lobe, which has not extended as far from the
cluster center as the upper lobe.  The radio emission shown in Figures \ref{fig:run1_radio} and \ref{fig:run2_radio} was contained within the boundaries of the X-ray cavities seen in
Figures \ref{fig:run1_xray} and \ref{fig:run1_xray} respectively.  

In addition to qualitative morphological evaluations that reveal simple dynamical relationships,  the above synthetic observations can be examined quantitatively to good effect, as well.  For example, \citet{treg02} used such
observations at multiple frequencies to test observational methods for measuring magnetic field strengths and properties 
of the CR population.  In addition to total intensity maps, our synthetic observation 
tool can also take advantage of the vector magnetic field  and CR electron plasma information in the simulations to construct polarization maps that include Faraday rotation.  These can be used, for example, to test techniques for measuring the structure of
magnetic fields in the AGN and in the ICM surrounding it.  We defer those studies to subsequent publications.


\section{The Role of Cluster ``Weather''}
\label{weather}
The asymmetries between the jet lobe and X-ray cavity pairs in these
simulations clearly demonstrate that asymmetries in the properties of the
ICM, that is, cluster ``weather patterns'', can strongly influence the morphology 
of radio lobes and ICM cavities.  
Since most radio galaxies in cluster environments show some degree of
distortion, it is useful to understand how to establish the dynamical links 
to observed AGN outflow distortions as metrics of the ICM structure.
In our jet simulations we can exclude direct gravitational forces as contributors to 
nonradial distortions of the AGN plasma and entrained ICM, since the gravitational 
potential in action during the simulations
was spherically symmetric. Nonspherical gravity was essential, of course, in the 
development of the ICM conditions present at the start of our simulations,
especially those we describe as sloshing.

A simple model can illuminate the main character of the AGN/ICM-weather
interactions at work during our simulations. For this purpose
we can picture the AGN outflows
as being ejected into a wind at some angle to the jet axis. It is 
sufficient for our purposes to look at the two limiting cases having the winds
aligned with the jet axis (i.e., ``head'' and ``tail'' winds)
and orthogonal to the jets (``cross'' winds).  For simplicity we assume
a uniform ICM and steady jets. In doing this we should remember that our simulated environment is
nonuniform and the simulated jets are intermittent. Still, the model can provide some
insights and consistency checks.

In the head and tail (aligned) wind case that applies approximately to our {\bf R2} simulation
we expect the length of the jet structure, $\ell$, on the tail wind
side ($v_w>0$ assuming $v_j>0$) to be extended in comparison to a stationary ICM, while the head wind
side ($v_w<0$) will be shortened. A modified version of the classic calculation
to balance the jet thrust against ram pressure from the ICM as the jet advances
\citep{bland74} provides a simple estimate of the wind's influence.  The model estimates,
if the local ICM density is $\rho_{ICM}$ and
the jet thrust is $F_j$, that the rate at which the jet ``head'' extends, $v_h > 0$, is
\begin{align}
v_h &\approx v_{h0} \left[1 + \frac{v_w}{|v_w|}\left(\frac{\rho_{ICM} v_w^2 \pi r_h^2}{F_j}\right)^{1/2}\right] \\ \nonumber
    &= v_{h0}\left(1 + \frac{v_w}{v_j}\sqrt{\chi}\right),
\end{align}
where $r_h$ is the cross sectional radius of the head, and
\begin{align}
v_{h0} = \frac{v_j}{1 + \sqrt{\chi}},
\end{align}
with $\chi = (r_h/r_j)^2 (\rho_{ICM}/\rho_j)$. In the applicable,``light jet''
limit, $\sqrt{\chi}>>1$ this gives the intuitively obvious result,
\begin{align}
v_h \approx v_{h_0} + v_w.
\end{align}
Setting the (observed) ratio of lengths, $\delta = \ell_+/\ell_- \sim (v_{h_0} + |v_w|)/(v_{h_0} - |v_w|)$, 
one can find, noting
that $v_{h0}$ is the average extension rate on the two sides, an estimate of
the wind speed as, $|v_w| \sim v_{h0} (\delta + 1)/(\delta - 1)$. 

At the end of our {\bf R2} simulation ($t \approx$ 200 Myr) one jet cavity extends roughly 300 kpc 
from the AGN, while the other reaches a distance only about 200 kpc
from the AGN (see Figure \ref{fig:run2_color}). Thus, we can estimate $\delta \sim 1.5$.
The average length is then about 250 kpc,
which over the approximately 200 Myr life of the AGN activity, corresponds to
an average extension rate ($\sim v_{h0}$) around 2500 km s$^{-1}$ (approximately Mach 5).
Using the relation derived in the previous paragraph this would produce an
estimate for the ICM wind speed, $v_w \sim$ 500 km s$^{-1}$. 
On these scales the actual ICM flow speed is around $400~{\rm km~s}^{-1}$ (see Figure \ref{fig:cluster_init}), so given the simplicity of the estimate,
consistent with this result.

The orthogonal, cross wind
case, related to our {\bf R1} simulation, is familiar in NAT and WAT contexts. In particular a transverse
ram pressure from the wind will deflect the jet and its cocoon \cite[e.g.,][]{begel79,jones79}. In the absence of a wind the jets will deposit
momentum, energy (including pressure)
and AGN plasma in the head region. The energy and plasma will flow backwards
and expand  in response to the added pressure to form a AGN plasma cocoon, which is represented by X-ray cavities in clusters. The cavities
are reasonably close to pressure balance with the surrounding ICM \citep[e.g.,][OJ10]{mcnamara07}, so the rate
of lateral expansion will be of the order of the ICM sound speed, $c_{ICM}$. The {\bf R1} cocoons at the end of the 
simulation have, on average, transverse radii, $R_c \sim$ 100 kpc, corresponding to mean expansion rates
around 500 km s$^{-1}$. That is close to $c_{ICM}$ and consistent with this expectation.

In a cross wind the cocoon plasma will experience a ram-pressure-based lateral force,
$F_w \sim 2\rho_{ICM} v_w^2 \ell R_c$, where $\ell$ is the length of the cocoon and $R_c$
is its transverse radius. The mass in the cocoon is roughly
$ \rho_c \ell \pi R_c^2$, with $\rho_c$ the density in the cocoon.
So, the time for $F_w$ to accelerate the cocoon to the wind speed is roughly,
$t_{wc} \sim (\pi/2) (\rho_c/\rho_{ICM}) (M_c/M_w) ~t_c$, where $M_w = v_w/c_{ICM}$
and $t_c \approx R_c/(M_c c_{ICM})$ is the time for the cocoon to expand. With both $M_c \sim 1$ and $M_w \sim 1$ the cocoon will
then reach terminal speed at rest in the ICM on a timescale less than it takes the coccon to expand.
On the other hand, if the wind speed exceeds the expansion speed; namely, if $M_w > M_c$, the cocoon will
be stripped from the jet, leaving a ``naked'' jet.  Our {\bf R1} jets do possess cocoons, since  $M_w <M_c$ .
Assuming that the cocoons move across the jet path at the wind speed,
we can estimate the down wind displacement angle of the cocoon compared to the jet direction at its source, $\theta_d$,  as $\tan{\theta_d} \sim v_w/v_h$. For the {\bf R1} structures
at the end of the simulation, this gives $\tan{\theta_d} \sim 0.15 - 0.2$, so $\theta_d \sim 10 - 15$ degrees. That is consistent with the deformations
visible on the left side of Figure \ref{fig:run1_color}.

For completeness we recall that a naked jet is deflected by a cross wind  over a characteristic length, $\ell_d$, \citep[e.g.][]{begel79,porter09}, 
\begin{align}
\ell_d(naked) \sim \frac{\rho_j}{\rho_{ICM}}\frac{v_j^2}{v_w^2} r_j \sim \frac{P_j}{P_{ICM}}\frac{M_j^2}{M_w^2} r_j.
\end{align}
The expression on the far right emphasizes that the deflection length, $\ell_d$ of a naked jet in pressure equilibrium
with its environment depends only on the jet radius and the relative Mach number in the jet flow to the Mach number of the wind. If the jet is cloaked it
will be deflected by a transverse pressure difference across the cocoon resulting
from $F_w$. The deflection length in this case can be estimated to be \citep{jones79},
\begin{align}
\ell_d(cloaked) \sim \frac{\rho_j}{\rho_{ICM}}\frac{v_j^2}{v_w^2} R_c = \frac{R_c}{r_j}\ell_d(naked),
\end{align}
so the bending will be more gradual for a given wind.  In our case the expected ratio is roughly an order of magnitude, consistent with the simulation
results.


\section{Conclusions}
\label{weather_conclusions}
We have presented two MHD simulations of AGN jets inside of a galaxy cluster that was adapted from an SPH cosmological
simulation to explore the effects of cluster ``weather'' on the jets and lobes.  The important results from this work are:

\begin{itemize}
\item Evidence that ICM flows in these simulations were sufficient to produce asymmetries between the opposing AGN jets and lobes
was presented.  Although pressure variations in the ICM (unrelated to the presence of a gravitational potential) and magnetic stresses were present, 
they were either not as large as the variations in the ICM motions or were very localized.  The existence of significant ``weather'' in the cluster
used for these simulations is interesting because it was specifically selected as a relatively relaxed system.  \S \ref{cluster_cond} showed that typical flow speeds
in the cluster were one the order of 400 km s$^{-1}$, which appear to be sufficent to modify the jets and lobes.  This result is in agreement with the 
predictions of \cite{hardcastle05}, who argued that flows as low as 100-300 km s$^{-1}$ could create bent WAT-like morphologies for jets similar 
to those in our model (high velocity and low density).  These results also show that the AGN outflows act as probes of bulk motions in the ICM, even 
in a seemingly relaxed cluster.  Simple dynamical models utilizing the influence of the ICM ram 
pressure on the  AGN cocoons seem to give useful
estimates of the ICM properties. This makes it possible to infer velocity structure information in environments where direct measurement is not possible.

\item Magnetic field lines anchored in the ICM are stretched outward by the intermittent jets.  This causes field lines to wrap around the jets and
AGN plasma in the lobes, and in localized regions, the stresses associated with the magnetic field were on the order or even greater than the
inertial stresses.  Such regions serve as a reminder that even for the ``high-$\beta$'' plasmas modeled here,
Maxwell stresses can significantly effect the evolution of a plasma and emphasize the importance of modeling these systems with MHD.

\item Synthetic X-ray observations of these simulations show complex cavity structures associated with the intermittent jets and ICM motions.
Multiple cavity pairs were ``observed'' as restarted jets pushed into cavities formed by previous outbursts.  In one example given, a new cavity
forming inside a larger cavity did not have an obvious companion cavity from the opposing jet.  Bright rims were seen outlining the cavities, and
they were associated with the injection of energy by restarting jets.  Additionally, cavities were seen misaligned with the jet launching region
due to the deflection of the jets by ICM flows.

\item Synthetic radio observations made during periods of jet activity resembled ``double-double'' radio galaxy sources with a bright pair of
jets interior to fainter sources further out in the cluster.  This state was temporary, however, as the cosmic ray electrons from the new outburst
would eventually reach the end of the outer lobe.  The ICM motions created a WAT morphology of the lobes in one simulation.
\end{itemize}

\begin{acknowledgments}
This work was supported at the University of Minnesota by NSF grant AST0908668. PJM was supported in part by
the Graduate Dissertation Fellowship at the University of Minnesota. KD acknowledge the supported 
by the DFG Priority Programme 1177, the DFG Cluster of Excellence 'Origin and Structure of the Universe' and
the DFG Research Unit 1254.  We thank the anonymous referee for help with improving the original manuscript. 
\end{acknowledgments}



\begin{thebibliography}{9}

\bibitem[Balsara \& Norman (1992)]{bals92}
Balsara, D. S. \& Norman, M. L. 1992, \apj, \textbf{393}, 631

\bibitem[Begelman \emph{et al.}(1979)]{begel79}
Begeleman, M. C., Rees, M. J. \& Blandford, R. D. 1979. Nature, \textbf{279}, 770

\bibitem[B\^{i}rzan \emph{et al.}(2004)]{birzan04}
B\^{i}rzan, L., Rafferty, D. A., McNamara, B. R., Wise, M. W., \& Nulsen, P. E. J. 2004, \apj, \textbf{607}, 800

\bibitem[B\^{i}rzan \emph{et al.}(2008)]{birzan08}
B\^{i}rzan, L., McNamara, B. R., Nulsen, P. E. J., Carilli, C. L., \& Wise, M. W. 2008, \apj, \textbf{686}, 859

\bibitem[Blandford \& Rees(1974)]{bland74}
Blandford, R. D. \& Rees, M. J. 1974, \mnras, \textbf{169}, 395

\bibitem[Blanton \emph{et al.}(2009)]{blanton09}
Blanton, E. L., Randall, S. W., Douglass, E. M., Sarazin, C. L., CLarke, T. E., \& McNamara, B. R. 2009, \apjl, \textbf{697}, L95

\bibitem[B\"{o}hringer \emph{et al.}(2010)]{bohringer10}
B\"{o}hringer, H., Pratt, G. W., Arnaud, M., Borgani, S., Croston, J. H., Ponman, T. J., Ameglio, S., Temple, R. F., 
\& Dolag, K. 2010, \aap, \textbf{514}, A32

\bibitem[Buote \& Tsai(1995)]{buote95}
Buote, D. A., \& Tsai, J. C. 1995, \apj, \textbf{553}, L15

\bibitem[Burns \emph{et al.}(1994)]{burns94}
Burns, J. O., Rhee, G., Owen, F. N., \& Pinkney, J. 1994, \apj, \textbf{423}, 94

\bibitem[Cassano \emph{et al.}(2010)]{cassano10}
Cassano, R., Ettori, S., Giacintucci, S., Brunetti, G., Markevitch, M., Venturi, T., \& Gitti, M. 2010, \apj, \textbf{721}, L82

\bibitem[Croston \emph{et al.}(2007)]{croston07}
Croston, J. H., Kraft, R. P., \& Hardcastle, M. J. 2007, \apj, \textbf{660}, 191

\bibitem[Diehl \emph{et al.}(2008)]{diehl08}
Diehl, S., Li, H., Fryer, C. L., Rafferty, D. 2008, \apj, \textbf{687}, 173

\bibitem[Dolag \& Stasyszyn(2009)]{mhd_gadget}
Dolag, K. \& Stasyszyn, F. 2009, \mnras, \textbf{398}, 1678

\bibitem[Dolag \emph{et al.}(2009)]{dolag09}
Dolag, K., Borgani, S., Murante, G., \& Springel, V. 2009, \mnras, \textbf{399}, 497

\bibitem[Douglass \emph{et al.}(2008)]{douglass08}
Douglass, E. M., Blanton, E. L., Clarke, T. E., Sarazin, C. L., Wise, M. 2008, \apj, \textbf{673}, 763

\bibitem[Dupke \& Bregman(2006)]{dupke06}
Dupke, R. A., \& Bregman, J. N. 2006, \apj, \textbf{639}, 781

\bibitem[Fabian \emph{et al.}(2003)]{fabian03}
Fabian, A. C., Sanders, J. S., Allen, S. W., Crawford, C. S., Iwasawa, K., Johnstone, R. M., Schmidt, R. W., \& Taylor, G. B. 2003, \mnras,
\textbf{344}, L43-L47

\bibitem[Fabian(1994)]{fabian94}
Fabian, A. C. 1994, \araa, \textbf{32}, 277

\bibitem[Fabian(1994)]{fabian04}
Fabian, A. C. 1994, \araa, \textbf{32}, 277

\bibitem[Hardcastle \emph{et al.}(2005)]{hardcastle05}
Hardcastle, M. J., Sakelliou, I., \& Worrall, D. M. 2005, \mnras, \textbf{359}, 1007

\bibitem[Heinz \emph{at al.}(2006)]{heinz06}
Heinz, S., Br\"{u}ggen, M., Young, A., \& Levesque, E. 2006, \mnras, \textbf{373}, L65

\bibitem[Hudson \emph{et al.}(2010)]{hudson10}
Hudson, D. S., Mittal, R., Reiprich, T. H., Nulsen, P. E. J., Andernach, H., \& Sarazin, C. L. 2010, \aap, \textbf{513}, A37

\bibitem[Ikebe \emph{et al.}(1996)]{ikebe96}
Ikebe, Y., Ezawa, H., Fukazawa, Y., Hirayama, M., Ishisaki, Y., Kikuchi, K., Kubo, H., Makishima, K., Matsushita, K., Ohashi, T., Takahashi, T., 
\& Tamura, T. 1996, \nat, \textbf{379}, 427

\bibitem[Jeltema \emph{et al.}(2005)]{jeltema05}
Jeltema, T. E., Canizares, C. R., Bautz, M. W., \& Buote, D. A. 2005, \apj, \textbf{624}, 606

\bibitem[Jones \& Kang(2005)]{jk05}
Jones, T. W., \& Kang, H., APh 2005, \textbf{24}, 75

\bibitem[Jones \emph{et al.}(1974)]{jos74}
Jones, T. W., O'Dell, S. L., \& Stein, W. A. 1974, \apj, \textbf{188}, 353

\bibitem[Jones \& Owen(1979)]{jones79}
Jones, T. W. \& Owen, F. N. 1979, \apj, \textbf{234}, 818

\bibitem[Joshi \emph{et al.}(2011)]{joshi11}
Joshi, S. A., Nandi, S., Saikia, D. J., Ishwara-Chandra, C. H., \& Konar, C. 2011, \mnras, \textbf{414}, 1397

\bibitem[Kraft \emph{et al.}(2003)]{kraft03}
Kraft, R. P., Va´zquez, S. E., Forman,W. R., Jones, C., Murray, S. S., Hardcastle, M. J., Worrall, D. M., \& Churazov, E. 2003, \apj, \textbf{592}, 129

\bibitem[Kraft \emph{et al.}(2007)]{kraft07}
Kraft, R. P., Nulsen, P. E. J., Birkinshaw, M., Worrall, D. M., Penna, R. F., Forman, W. R., Hardcastle, M. J., Jones, C., \& Murray, S. S. 2007, 
\apj, \textbf{665}, 1129

\bibitem[Laing \emph{et al.}(2008)]{laing08}
Laing, R. A., Bridle, A. H., Parma, P., Feretti, L., Giovannini, G., Murgia, M., \& Perley, R. A. 2008, \mnras, \textbf{386}, 657

\bibitem[Loken \emph{et al.}(1995)]{loken95}
Loken, C., Roettiger, K., Burns, J. O., \& Norman, M. 1995, \apj, \textbf{445}, 80

\bibitem[Markevitch \& Vikhlinin(2007)]{markevitch07}
Markevitch, M., \& Vikhlinin, A. 2007, Phys. Rep., \textbf{443}, 1

\bibitem[McNamara \emph{et al.}(2005)]{mcnamara05}
McNamara, B. R., Nulsen, P. E. J., Wise, M. W., Rafferty, D. A., Carilli, C., Sarazin, C. L., \& Blanton, E. L. 2005, \nat, \textbf{433}, 45

\bibitem[McNamara \& Nulsen(2007)]{mcnamara07}
McNamara, B. R., \& Nulsen, P. E. J. 2007, \araa, \textbf{45}, 117

\bibitem[Mendygral \emph{et al.}(2011)]{mend11a}
Mendygral, P. J., O'Neill, S. M. \& Jones, T. W. 2011, \apj, \textbf{730},100

\bibitem[Mendygral \emph{et al.}(2012a)]{mend11b}
Mendygral, P. J., Porter, D. H., Edmon, P. P., Delgado, J., Wesson, A. \& Jones, T. W. 2011 \apjs  (Mendygral et al. 2012a, to be submitted)

\bibitem[Mendygral \emph{et al.}(2012b)]{mend11c}
Mendygral, P. J., Delgado, \& Jones, T. W. 2011 \apjs  (Mendygral et al. 2012b,to be submitted)

\bibitem[Mitchell \emph{et al.}(2009)]{mitchell09}
Mitchell, N. L., McCarthy, I. G., Bower, R. G., Theuns, T., \& Crain, R. A. 2009, \mnras, \textbf{395}, 180

\bibitem[Morsony \emph{et al.}(2010)]{morsony10}
Morsony, B. J., Heinz, S., Bru\"{u}ggen, M., Ruszkowski 2010, \mnras, \textbf{407}, 1277

\bibitem[Nozawa \emph{et al.}(1998)]{nozawa}
Nozawa, S., Itoh, N., \& Kohyama, Y. 1998, \apj, \textbf{507}, 530

\bibitem[O'Neill \& Jones(2010)]{oj}
O'Neill, S. M., \& Jones, T. W. 2010, \apj, \textbf{710}, 180 (OJ10)

\bibitem[Owen \& Rudnick(1976)]{owen76}
Owen, F. N., \& Rudnick, L. 1976, \apjl, \textbf{203}, L107

\bibitem[Peterson \& Fabian(2006)]{peterson06}
Peterson, J. R., \& Fabian, A. C. 2006, \physrep, \textbf{427}, 1

\bibitem[Porter \emph{et al.}(2009)]{porter09}
Porter, D. H., Mendygral, P. J. \& Jones, T. W. 2009, AIP Conference Proc. 1201,
The Monster's Fiery Breath: Feedback in Galaxies, Groups and Clusters,
ed. S. Heinz \& E. Wilcots, New York: AIP, 259

\bibitem[Roettiger \emph{et al.}(1996)]{roettiger96}
Roettiger, K., Burns, J. O., \& Loken, C. 1996, \apj, \textbf{473}, 651

\bibitem[Ricker \& Sarazin(2001)]{ricker01}
Ricker, P. M., \& Sarazin, C. L. 2001, \apj, \textbf{561}, 621

\bibitem[Ryu \& Jones(1995a)]{rj95}
Ryu, D., \& Jones, T. W. 1995, \apj, \textbf{442}, 228

\bibitem[Ryu \emph{et al.}(1998)]{rj98}
Ryu, D., Miniati, F., Jones, T. W., \& Frank, A. 1998, \apj, \textbf{509}, 244

\bibitem[Santos \emph{et al.}(2008)]{santos08}
Santos, J.~S., Rosati, P., Tozzi, P., B{\"o}hringer, H., Ettori, S., \& Bignamini, A. 2008, \aap, \textbf{483}, 35

\bibitem[Schoenmakers \emph{et al.}(2000)]{schoenmakers00}
Schoenmakers, A. P., de Bruyn, A. G., R\"{o}ttgering, H. J. A., van der Laan, H., \& Kaiser, C. R., 2000, \mnras, \textbf{315}, 371

\bibitem[Tormen \emph{et al.}(1997)]{tormen97}
{Tormen}, G. and {Bouchet}, F.~R. and {White}, S.~D.~M. 1997, \mnras, \textbf{286}, 865

\bibitem[Tregillis \emph{et al.}(2001)]{treg01}
Tregillis, I., Jones, T. W., \& Ryu, D. 2001, \apj, \textbf{557}, 475

\bibitem[Tregillis \emph{et al.}(2002)]{treg02}
Tregillis, I. L., Jones, T. W., \& Ryu, D. 2002,  ASP Conference Proc. 250, Particles and Fields in Radio Galaxies, ed. RA
Laing \& KM Blundell, San Francisco: ASP, 336

\bibitem[Wise \emph{et al.}(2007)]{wise07}
Wise, M. W., McNamara, B. R., Nulsen, P. E. J., Houck, J. C., \& David, L. P. 2007, \apj, \textbf{659}, 1153

\bibitem[Ventimiglia \emph{et al.}(2008)]{ventimiglia08}
Ventimiglia, D. A., Voit, G. M., Donahue, M., Ameglio, S. 2008, \apj, \textbf{685}, 118

\bibitem[White(1996)]{1996clss.conf..349W}
White, S. M. D. 1996, in Cosmology and Large Scale Structure, ed. R. Schaeffer, J Silk, M Spiro, \& J Zinn-Justin, Amsterdam: Elsevier, 349

\bibitem[Yoshida \emph{et al.}(2001)]{2001MNRAS.328..669Y}
{Yoshida}, N. and {Sheth}, R.~K. and {Diaferio}, A. 2001, \mnras, \textbf{328}, 669

\bibitem[Zhuravleva \emph{et al.}(2011)]{zhura11}
Zhuravleva, I., Churazov, E., Sazonov, S., Sunyaev, R. \& Dolag, K. 2011, Ast. Lett., \textbf{37}, 141

\bibitem[Zuhone \emph{et al.}(2010)]{zuhone10}
Zuhone, J. A., Markevitch, M. \& Johnson, R. E. 2010, \apj, 717, 908
\end{thebibliography}
\end{document}